\documentclass[twocolumn,aps,pra,showpacs,raggedbottom,nobalancelastpage,amssymb,superscriptaddress,longbibliography]{revtex4-1}

\usepackage{graphicx}
\usepackage{amsmath}
\usepackage{amssymb}
\usepackage{bm}
\usepackage[usenames]{color}
\usepackage{amsfonts}
\usepackage{appendix}

\usepackage{comment}

\usepackage[colorlinks=true , citecolor=blue,urlcolor=blue]{hyperref}

\newcommand{\ie}{{i.e.}}

\newcommand{\e}{\textrm{e}}

\newcommand{\nbf}{\mathbf{n}}
\newcommand{\qbf}{\mathbf{q}}
\newcommand{\qp}{\mathbf{q}_{\perp}}
\newcommand{\pp}{\mathbf{p}_{\perp}}
\newcommand{\rp}{\mathbf{r}_{\perp}}
\newcommand{\ah}{\hat{a}}
\newcommand{\ahd}{\hat{a}^{\dagger}}
\newcommand{\Eav}{E_{\mathrm{av}}}
\newcommand{\epsav}{\varepsilon_{\mathrm{av}}}
\newcommand{\Jeff}{J_{\mathrm{eff}}}

\newcommand{\Jnature}{Nature (London)}
\newcommand{\Jnatphys}{Nature Phys.}

\newcommand{\Jscience}{Science}

\newcommand{\Jprl}{Phys. Rev. Lett.}
\newcommand{\Jpr}{Phys. Rev.}
\newcommand{\Jpra}{Phys. Rev. A}
\newcommand{\Jprb}{Phys. Rev. B}

\newcommand{\Jprx}{Phys. Rev. X}

\newcommand{\Jepl}{Europhys. Lett.}
\newcommand{\Jnjp}{New J. Phys.}

\newcommand{\JRepProgPhys}{Rep. Prog. Phys.}

\newcommand{\Jadvphys}{Adv. Phys.}

\definecolor{Nathanpurple}{rgb}{0.5,0.,0.5}

\definecolor{Nathanblue}{rgb}{0.96,0.24,0.00}

\definecolor{Nathanred}{rgb}{0.06,0.24,0.90}

\def\be{\begin{equation}}
\def\ee{\end{equation}}

\begin{document}
\title{Parametric Instabilities in Resonantly-Driven Bose-Einstein Condensates}
\author{S. Lellouch}
\email[]{samuel.lellouch@univ-lille1.fr}
\affiliation{Center for Nonlinear Phenomena and Complex Systems, Universit\'e Libre de Bruxelles, CP 231, Campus Plaine, B-1050 Brussels, Belgium}
\affiliation{Laboratoire de Physique des Lasers, Atomes et Molécules,
Université Lille 1 Sciences et Technologies, CNRS ; F-59655 Villeneuve d’Ascq Cedex, France}
\author{N. Goldman}
\email[]{ngoldman@ulb.ac.be}
\affiliation{Center for Nonlinear Phenomena and Complex Systems, Universit\'e Libre de Bruxelles, CP 231, Campus Plaine, B-1050 Brussels, Belgium}
\begin{abstract}
Shaking optical lattices in a resonant manner offers an efficient and versatile method to devise artificial gauge fields and topological band structures for ultracold atomic gases. This was recently demonstrated through the experimental realization of the Harper-Hofstadter model, which combined optical superlattices and resonant time-modulations. Adding inter-particle interactions to these engineered band systems is expected to lead to strongly-correlated states with topological features, such as fractional Chern insulators. However, the interplay between interactions and external time-periodic drives typically triggers violent instabilities and uncontrollable heating, hence potentially ruling out the possibility of accessing such intriguing states of matter in experiments. In this work, we study the early-stage parametric instabilities that occur in systems of resonantly-driven Bose-Einstein condensates in optical lattices. We apply and extend an approach based on Bogoliubov theory~[PRX 7, 021015 (2017)] to a variety of resonantly-driven band models, from a simple shaken Wannier-Stark ladder to the more intriguing driven-induced Harper-Hofstadter model. In particular, we provide \textit{ab initio} numerical and analytical predictions for the stability properties of these topical models. This work sheds light on general features that could guide current experiments to stable regimes of operation.
\end{abstract}

\date{\today}
\maketitle

\section{Introduction}

	Driving quantum systems periodically in time has been proposed as a versatile tool to generate unusual quantum phases of matter~\cite{oka_09,kitagawa2010,lindner2011,cayssol2013,mitrano2016possible,aidelsburger2017artificial}. In the context of ultracold quantum gases, it was shown that subjecting neutral atoms to an external time-periodic drive could be used to design artificial gauge fields in these systems~\cite{goldman2014a,goldman2014b,bukov2014,Eckardt:2016Review,aidelsburger2017artificial}, hence opening promising perspectives in the quantum simulation of topological states of matter~\cite{aidelsburger2014,jotzu2014,Goldman:2016Review} and quantum magnetism~\cite{eckardt2010,struck2011,Eckardt:2016Review}; see also the recent work~\cite{Gorg_2017}. The underlying concept of \textit{Floquet engineering}~\cite{cayssol2013,goldman2014b,bukov2014,Goldman:2016Review,Eckardt:2016Review} builds on the fact that the dynamics of periodically-driven systems can be well described by a \textit{static} effective Hamiltonian, whose properties can be suitably designed by tailoring the driving protocol. On the experimental side, this idea has been extensively used to design artificial gauge fields for neutral atoms~\cite{aidelsburger2011,aidelsburger2013,miyake2013,aidelsburger2014,jotzu2014,kennedy2015,tai2017microscopy,tarnowski2017characterizing}, some of which lead to non-trivial geometrical and topological properties~\cite{Goldman:2016Review}.

	A particularly interesting class of periodically-driven setups is that featuring \textit{resonant} time-modulations~\cite{kolovsky_11,bermudez2011synthetic,hauke_12,goldman2015a,creffield2016,price2017synthetic}, in which the driving frequency $\omega$ resonates with an energy separation $\Delta\!\approx\!\hbar \omega$ that is inherent to the underlying static system. In particular, such schemes can be exploited to finely control the tunneling matrix elements connecting neighboring sites of a lattice, and can be simply implemented by resonantly modulating a superlattice or a Wannier-Stark ladder; see Refs.~\cite{sias2008,mukherjee2015modulation} for experimental realizations and Ref.~\cite{Eckardt:2016Review} for a review. This so-called \emph{photon-assisted tunneling} effect constitutes a natural ingredient for the generation of artificial fluxes within cold-atom systems~\cite{kolovsky_11,hauke_12,goldman2015a,creffield2016}, as exemplified by the recent experimental realizations of the Harper-Hofstadter model~\cite{aidelsburger2011,aidelsburger2013,miyake2013,kennedy2015,aidelsburger_14,tai2017microscopy}, a 2D lattice penetrated by a uniform magnetic flux~\cite{hofstadter1976}; the latter band model is of particular interest, due to its rich topological band structure~\cite{kohmoto1989zero,moller2015fractional}. 
	
	However, a major challenge remains in this context, namely, the addition of inter-particle interactions in view of creating novel strongly-correlated states of matter,~e.g.~fractional Chern insulators~\cite{regnault2011fractional,sterdyniak2013series,bergholtz2013topological,grushin_14,moller2015fractional}. Indeed, the interplay between inter-particle interactions and an external drive, such as lattice shaking, has been shown to lead to significant heating and losses in experiments involving Bose-Einstein condensates (BEC)~\cite{aidelsburger_14,reitter2017}. The complexity of this issue lies on the fact that several underlying mechanisms are believed to be responsible for those undesired effects, and these possibly interplay in a complex manner, as we now explain. 
	
	On the one hand, at very short times, time-modulated BECs are believed to be mostly affected by strong parametric instabilities, which are characterized by an exponential growth of collective (Bogoliubov) excitations and are accompanied with a fast decay of the BEC; such processes were thoroughly characterized in our previous work~\cite{lellouch2017}, where instability rates, stability diagrams and robust physical signatures of such processes were obtained for simple non-resonant shaken systems. Other approaches have been adopted to characterize dynamical instabilities in shaken BECs~\cite{modugno2004,Kramer:2005,tozzo2005,creffield2009,bukov2015}, and a recent study even pointed out the possibility of dynamically stabilizing a modulated BEC, inspired by the Kapitza pendulum~\cite{martin2017}; the experimental evidence of staggered-states in time-modulated BECs, whose formation also stems from an instability involving the external drive and collective excitations, was recently reported in Ref.~\cite{michon2017}; we note that time-modulating the trapping potential can also be exploited to create correlated excitations in BECs~\cite{jaskula2012acoustic}. 
	
	On the other hand, at longer times, dissipative processes are expected to be dominated by scattering events, which are typically associated with two-body processes and are captured by the so-called Floquet-Fermi golden rule~\cite{bilitewski_14,bilitewski2015,aidelsburger2014}. Finally, the role of (resonant) inter-band transitions~\cite{weinberg2015,strater2016,reitter2017,Quelle2017}, and the formation of collective emissions of matter-wave jets upon driving~\cite{clark2017}, have been analyzed in very recent ultracold-atom experiments. 

	The aim of this paper is to analyze the onset of parametric instabilities in resonantly-modulated BEC's. Building on the tools developed in Ref.~\cite{lellouch2017}, we identify the general features of those instabilities that occur in the resonant-modulations context, and provide useful results that could be readily applied to topical systems, such as the driven-induced Harper-Hofstadter model~\cite{aidelsburger2011,aidelsburger2013,miyake2013,kennedy2015,aidelsburger_14,tai2017microscopy}. 
	
	The rest of the paper is organized as follows:~In Sec.~\ref{sec:GenMeth}, we recall the general method of Ref.~\cite{lellouch2017} to treat parametric instabilities. In Sec.~\ref{sec:1DRes}, we apply this method to a simple resonantly-shaken Wannier-Stark ladder, highlighting the main features of these driven-induced instabilities. We then address the topical case of an optical lattice that is modulated by a secondary \emph{moving} lattice, and which includes a space-dependent phase~\cite{miyake2013,kennedy2015,aidelsburger2013}; we first study the 1D case in Sec.~\ref{sec:1Dfluxpi}, and then discuss the 2D configuration (leading to the Harper-Hofstadter Hamiltonian~\cite{miyake2013,kennedy2015,aidelsburger2013}) in Sec.~\ref{sec:2DHof}. Final remarks are provided in the concluding section~\ref{conclusion}.

	\section{General method}
	\label{sec:GenMeth}
	
	We first briefly summarize the general method of Ref.~\cite{lellouch2017} to study parametric instabilities in periodically-driven BEC lattice systems; we point out that similar or complementary approaches have been proposed to characterize dynamical instabilities in Refs.~\cite{modugno2004,Kramer:2005,tozzo2005,creffield2009,bukov2015,salerno2016}.
	
\subsubsection*{Linear stability analysis} 

	Consider a weakly-interacting Bose gas, in a generic periodically-driven system of period $T\!=\!2\pi/\omega$;~this discussion disregards the system dimension, and/or the presence of a lattice. To study the stability properties of a potential BEC, the general idea is to assume that the system is initially fully-condensed in some well-defined state and to perform a linear stability analysis around this specific state.  Let us denote by $a_{\nbf}^{(0)}(t\!=\!0)$ the condensate wavefunction at initial time $t\!=\!0$, where $\nbf$ is a generic (possibly multidimensional, discrete or continuous) index for spatial position. This state could be set by hand (based on analytical arguments), or it could be more precisely estimated for a given experimental protocol, by numerically implementing the full preparation sequence  or using Floquet adiabatic perturbation theory~\cite{pweinberg_15,novicenko2016}.\\ 
	
	Given this state $a_{\nbf}^{(0)}(t=0)$ as an input, we are interested in the stability of the full time-dependent solution $a_{\nbf}^{(0)}(t)$. This property can be evaluated by following the guideline below:

	(i) In the weakly-interacting regime, this solution is governed by the time-dependent Gross-Pitaevskii Equation (tGPE), whose exact form depends on the precise model under consideration. Thus, we first determine the time-evolution of the condensate wavefunction $a_{\nbf}^{(0)}(t)$, by solving the (tGPE) with the initial condition $a_{\nbf}^{(0)}(t=0)$. This can be performed numerically using direct real-time propagation in real space. We stress that this calculation is based on the full time-dependent equations, so that the dynamics of the BEC is exactly computed (including all micromotion effects~\cite{goldman2014b}).

	(ii) Given the time-dependent solution for the condensate wave function $a_{\nbf}^{(0)}(t)$, we analyse its stability by considering a small perturbation 
	\begin{equation}
	a_{\nbf}(t)\!=\!a_{\nbf}^{(0)}(t)[1\!+\!\delta a_{\nbf}(t)],
	\label{eq:delta_a}
	\end{equation} 
and linearizing the (tGPE) in $\delta a_{\nbf}$.  
This yields the time-dependent Bogoliubov-de Gennes equations, which take the general form
\begin{equation}
i \left( \begin{matrix} \dot{\delta a_{\nbf}}  \\ \dot{\delta a}_{\nbf}^{*} \end{matrix} \right) =\mathcal{L}(t) \left( \begin{matrix} \delta a_{\nbf}  \\ \delta a_{\nbf}^{*} \end{matrix} \right),
\label{eq:BdGE}
\end{equation}
where $\mathcal{L}(t)$ is a $T$-periodic operator (we set $\hbar\!=\!1$ here and in the following). 
Based on this time-periodicity, it is convenient to exploit the Floquet theorem and to focus the stability analysis on the ``time-evolution" (propagator) matrix $\Phi(T)$, which is obtained by time-evolving Eq.~(\ref{eq:BdGE}) over a single period $T$. From the knowledge of $\Phi(T)$, we extract the ``Lyapunov" exponents $\epsilon_{\qbf}$, which are related to the eigenvalues $\lambda_{\qbf}$ of $\Phi(T)$ through the relation $\lambda_{\qbf}=\e^{-i\epsilon_{\qbf} T}$; here we explicitly introduced the momentum ${\qbf}$, which will be used to index the corresponding excitation modes. The appearance of Lyapunov exponents with positive imaginary parts indicates a dynamical instability~\cite{tozzo2005,creffield2009}, \ie~an exponential growth of the corresponding modes, given by the rate $s_{\qbf}=\mathrm{Im}\; \epsilon_{\qbf}$.

\subsubsection*{Observables}

As previously discussed in Ref.~\cite{lellouch2017}, a first quantitative indicator of the instability is the maximum growth rate of the spectrum, 
	\begin{equation}
		\Gamma\equiv \max_{\qbf} s_{\qbf},
		\label{eq:GammaDef}
	\end{equation}
which, in the following, will be referred to as the \textit{instability rate} $\Gamma$. 
This instability rate is independent of the reference frame (or gauge). It quantifies the parametric instabilities occurring in the system in the sense that it governs the stroboscopic dynamics ($t\!=\!T \times \text{integer}$) of physical observables in the system. For instance, $\Gamma>0$ indicates that physical observables (e.g.~the total energy, the depleted fraction, ...) will exponentially grow up with the rate $2\Gamma$~\footnote{The factor of $2$ stems from the fact that $\Gamma$ is defined in amplitude while physical observables involve squared moduli of wavefunctions.}. \\
	
	Another relevant indicator is the most unstable mode
	\begin{equation}
		\qbf^\mathrm{mum}\equiv \mathrm{argmax}_{\qbf} s_\qbf.
		\label{eq:qmumDef}
	\end{equation}
The excitations momentum distribution indeed shows a pronounced contribution around $\qbf^\mathrm{mum}$ and structures of momentum $\qbf^\mathrm{mum}$ develop in real/momentum space, producing clear signatures of those parametric instabilities. We stress that, due to the factorization of the BEC wavefunction in Eq.~\eqref{eq:delta_a}, all momentum modes $\qbf$ (and hence, the most unstable mode) are always defined relatively to the ground-state (which may not always be associated with a vanishing momentum). \\
	
As pointed out in Ref.~\cite{lellouch2017}, saturation effects generically alter the instability rates ($\Gamma$) predicted by Bogoliubov theory at longer times; these effects are mainly attributed to couplings between the Bogoliubov modes. In this sense, the parametric instabilities investigated in this framework are truly associated with short-time dynamics. While the instability rate must therefore be treated as a dynamical quantity (affected by saturation effects and other, possibly incoherent, mechanisms~\cite{bilitewski_14,bilitewski2015}), the most unstable mode $\qbf^\mathrm{mum}$ and the associated structures developing in real and momentum space (e.g.~in the momentum distribution) are found to be very robust; this could, therefore, provide clear signatures of parametric instabilities in realistic experimental configurations.

	\section{A first example: The resonantly-shaken Wannier-Stark ladder}
	\label{sec:1DRes}

We first apply this method to a simple one-dimensional (1D) resonantly-driven Wannier-Stark ladder; we note that this toy-model is a direct extension of the shaken 1D lattice studied in Ref.~\cite{lellouch2017}. Beyond its physical interest, this Section aims at demonstrating how the (numerical and analytical) tools that were previously developed can indeed be successfully applied in the context of resonantly-driven models.

\subsection{The model}

\qquad We consider a system of weakly-interacting bosons, trapped in a shaken 1D Wannier-Stark ladder, as described by the periodically-driven Bose-Hubbard Hamiltonian~\cite{Eckardt:2016Review}
 \begin{eqnarray}
  \hat{H}(t)& =&-J\sum_n (\ahd_{n+1}\ah_n+ \mathrm{h.c.})+ \frac{U}{2} \sum_n \ahd_n\ahd_n\ah_n\ah_n \nonumber\\
 & & + \sum_n \Big[\Delta n \ahd_{n}\ah_{n} + K\cos(\omega t) n \ahd_n\ah_n \Big],
 \label{eq:H}
 \end{eqnarray}
where $\ah_{n}$ annihilates a particle at lattice site $n$, $J>0$ denotes the tunneling amplitude of nearest-neighbor hopping, and $U>0$ is the repulsive on-site interaction strength. The second line in Eq.~\eqref{eq:H} captures the on-site potential term, which contains two effects: the Wannier-Stark-ladder potential, which introduces an energy shift $\Delta>0$ between consecutive sites, and a time-periodic modulation of amplitude $K$ and frequency $\omega=2\pi/T$; this time-modulation simply corresponds to an external shaking of the 1D optical lattice, as viewed from the moving frame~\cite{Eckardt:2016Review,creffield2016}. The frequency modulation $\omega$ is chosen to be resonant with the offset $\Delta$, i.e. $\Delta=l\omega$ with $l$ denoting some integer; see Refs.~\cite{sias2008,mukherjee2015modulation}.

It is well known that the main effect of the time-modulation is to restore the tunneling, which is suppressed by the strong offset $\Delta$; this is reflected in the (non-interacting) Floquet effective Hamiltonian associated with Eq.~(\ref{eq:H}), and which reads~\cite{sias2008,choudhury2015a,goldman2015a,Eckardt:2016Review}
 \begin{equation}
 \hat{H}_{\mathrm{eff}}^{(0)}= - \Jeff \sum_{n}  [\ahd_{n+1}\ah_{n}+\mathrm{h.c.}],
 \label{eq:Heff}
 \end{equation}
where the effective tunneling is $\Jeff=J\mathcal{J}_{-l}(K/\omega)$, with $\mathcal{J}_{l}$ the l-th order Bessel function, and where we set $U\!=\!0$.

\subsection{Numerical results}

To compute the instability rates of the system, we first numerically implement the procedure detailed in Sec.~\ref{sec:GenMeth}.
The initial state $a^{(0)}_n(t=0)$ is chosen  to be the ground state of the naive ``interacting effective Hamiltonian"
 \begin{equation}
 \hat{H}_{\mathrm{eff}}^{(0)}+\dfrac{U}{2} \sum_{n}\ahd_{n}\ahd_{n}\ah_{n}\ah_{n},
 \label{eq:Heff2}
 \end{equation}
which is a reasonable approximate for an experimentally-prepared ground-state; in general, we note that this ground-state can be numerically determined using imaginary time propagation. For the specific model under consideration, we find that $a_n^{(0)}(t=0)$ is the Bloch state $e^{ip_0n}$ of momentum $p_0\!=\!0$ if $\Jeff\!>\!0$ (homogeneous condensate), and $p_0\!=\!\pi$ for $\Jeff\!<\!0$: see analytical details in Appendix A. This fact is reminiscent of the non-resonant case; see Refs.~\cite{creffield2009,lellouch2017}.

		\begin{figure}[!t]
			\begin{center}
				\includegraphics[width=8cm]{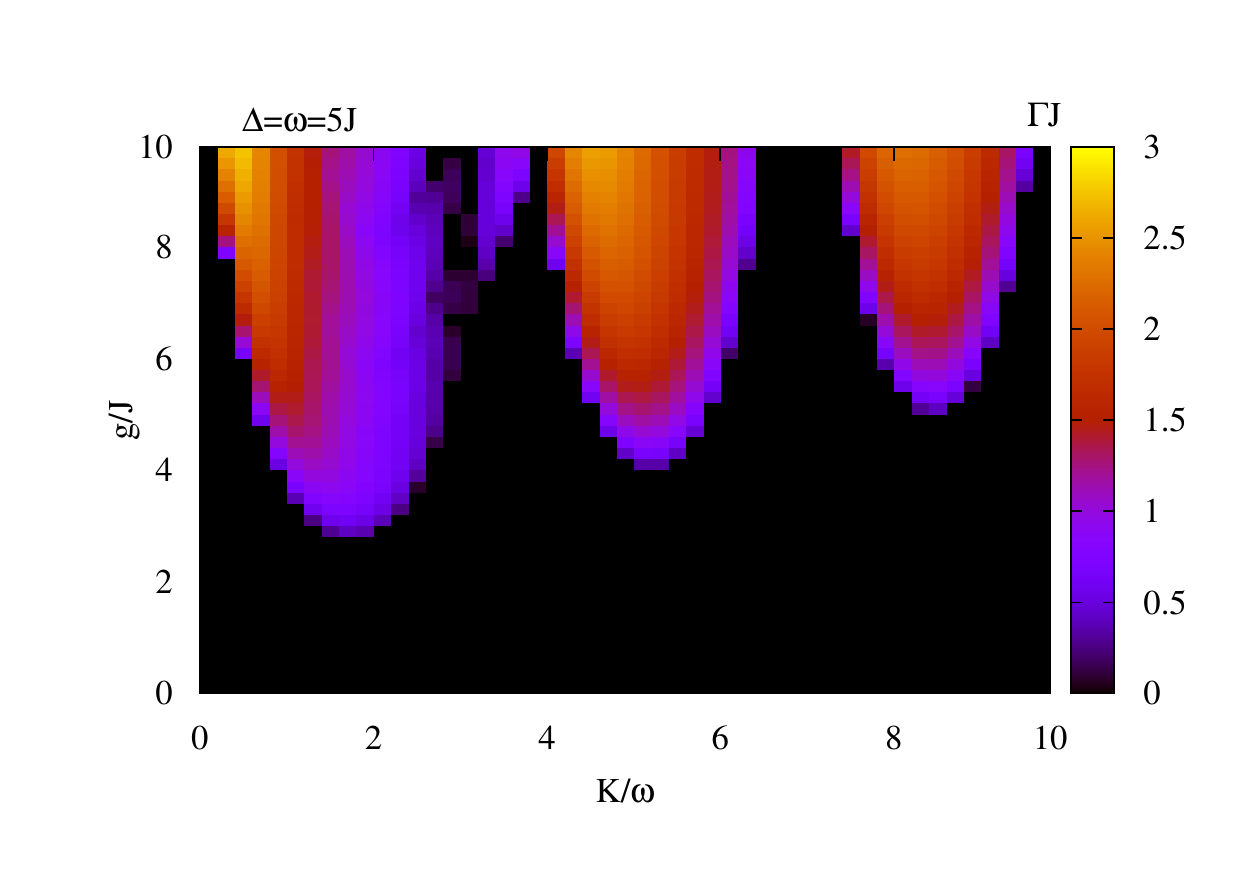}
				\includegraphics[width=8cm]{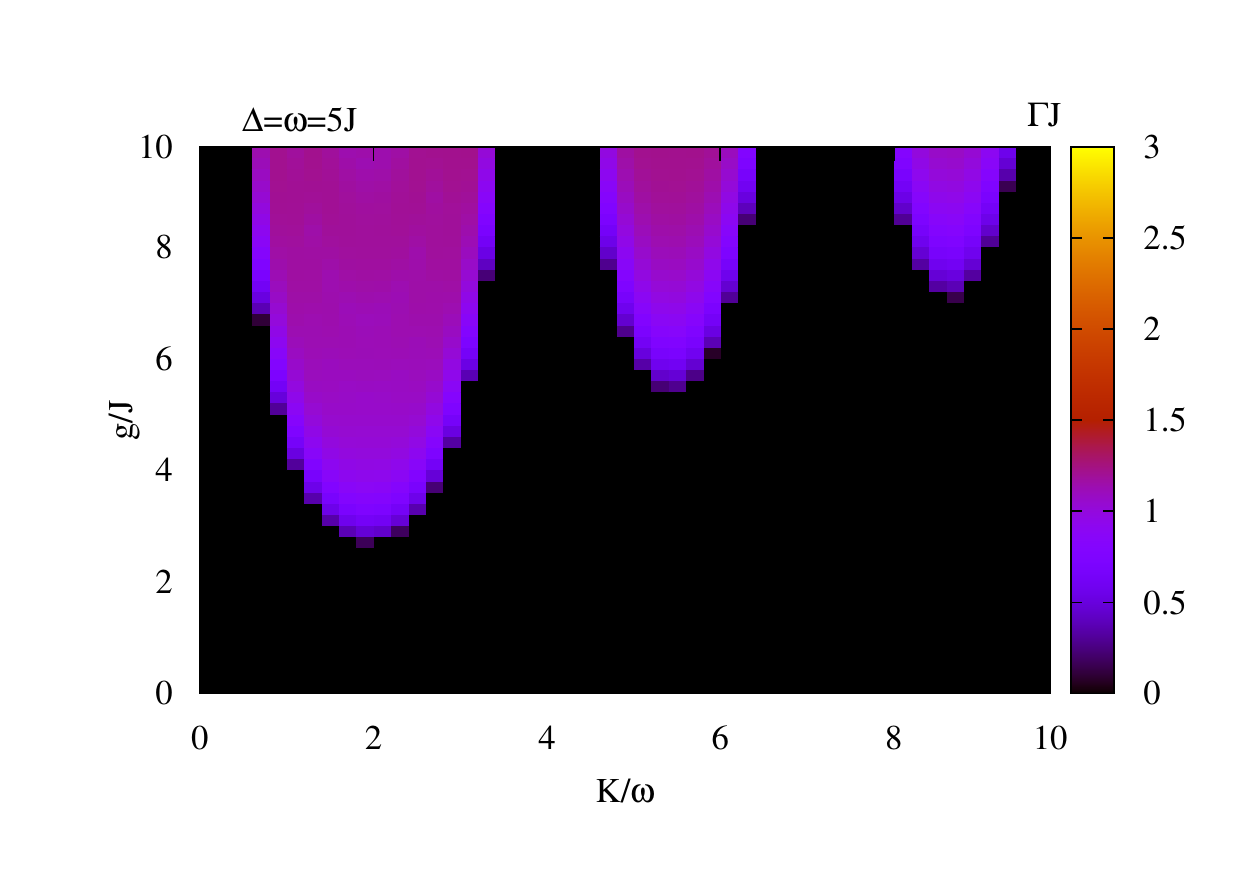}
			\end{center}
			\caption{Stability of the resonantly-driven Wannier-Stark ladder. Numerical (top) and analytical (bottom) instability rate $\Gamma$ as a function of the interaction strength $g=U\rho$ and the modulation amplitude $K/\omega$. Here we set $\omega=\Delta=5J$.
				\label{fig:NumRes}
			}
		\end{figure}

		Figure~\ref{fig:NumRes} displays the behavior of the instability rate $\Gamma$ as a function of the interaction strength $g\!=\!U\rho$ (with $\rho$ the condensate density, which enters the normalization of the initial state) and modulation amplitude $K/\omega$, in the purely resonant case ($l=1$) and for a driving frequency $\omega\!=\!5J$.
We note that the general features are very similar to what was observed in Refs.~\cite{creffield2009,lellouch2017} in the non-resonant case : the stability diagram displays lobes, which are separated by stable regions corresponding to cancellation points of the effective tunneling $\Jeff$ (here the zeros of the function $\mathcal{J}_{-1}(K/\omega)$). For large enough driving frequencies (such as in Fig.~\ref{fig:NumRes}), the system is stable at low values of $g$, and a transition to instability appears at finite $g$; close to the transition, these instabilities are found to be dominated by the Bogoliubov mode $q^{\text{mum}}\!=\!\pi$, which is the most unstable one; for smaller values of $\omega$ (smaller than the effective free-particle bandwidth, i.e.~$\omega<4|\Jeff|$ here), the system would be unstable for any non-zero interaction strength, with the most unstable mode corresponding to $q^{\text{mum}}\!<\!\pi$~\cite{lellouch2017}.\\

We find that this situation is very general: For other values of $l$, we find similar stability diagrams, except that the positions of the lobes are now governed by the Bessel function $\mathcal{J}_{-l}(K/\omega)$ (instead of $\mathcal{J}_{-1}$); besides, similar conclusions hold for the nature of the most unstable mode.
		
	\subsection{Analytical approach}\label{section_analytical}
	
	The numerical results described in the previous Section can be understood using the analytical method developed in Ref.~\cite{lellouch2017}, which can indeed be readily transposed to the present model; see Appendix A for the full calculations.
	
	The main idea is that the Bogoliubov equations of motion~(\ref{eq:BdGE}) can be mapped onto a parametric oscillator model~\cite{landau1969,bukov2015}, a seminal model of periodically-driven harmonic oscillator known to display dynamical instabilities as soon as the drive frequency approaches twice its own (intrinsic) frequency.
	Similarly here, we find (see Appendix A) that each Bogoliubov mode $q$ will display a dynamical instability (characterized by an exponential growth of its population) whenever the drive frequency $\omega$ approaches its \textit{time-averaged} Bogoliubov energy, namely, $\omega\!\approx\!\Eav(q)$, with
	\begin{align}
		&\Eav(q)=\sqrt{4|\Jeff|\sin^2(q/2)(4|\Jeff|\sin^2(q/2)+2g)}.
		\label{eq:Eav}
	\end{align}
Note that the latter represents the Bogoliubov dispersion associated with the linearized GPE, based on the naive ``interacting effective Hamiltonian" in Eq.~(\ref{eq:Heff2}).
	As detailed in Appendix A, the growth rate $s_q$ associated with this instability can  be computed analytically using a perturbative method~\cite{landau1969,lellouch2017}. From the knowledge of those individual rates, it is then straightforward to infer the total instability rate $\Gamma$ and the most unstable mode [see Eqs.~(\ref{eq:GammaDef}) and~(\ref{eq:qmumDef})].

	Figure~\ref{fig:NumRes} (bottom panel) shows the analytical stability diagram associated with the present model, as obtained from the aforementioned analytical perturbative method; as a technical remark, we point out that the latter calculation was performed up to second order with respect to the perturbation's amplitude $\alpha_q$ defined in Eq.~\ref{eq:appalpha}; see Appendix A and Refs.~\cite{landau1969,lellouch2017} for details. It shows very good agreement with the numerical diagram of Fig.~\ref{fig:NumRes}, and we attribute the small discrepancies to the perturbative nature of the method (higher order terms are generically expected to provide small corrections).
	
	Importantly, the analytical approach explains the existence of stable regions in the vicinity of the cancellation points $\Jeff\!\approx\!0$, as identified in Ref.~\cite{creffield2009}. Indeed, when $\Jeff$ vanishes, the time-averaged Bogoliubov dispersion $\Eav(q)$ becomes trivially flat, so that no excitation mode fulfills the resonance criterion $\Eav(q)\approx\omega$ associated with the existence of parametric instability.
	
Besides, the analytical approach also offers a simple view on the boundaries separating stable and unstable regions; in particular, it predicts the nature of the most unstable mode responsible for the onset of instabilities in the vicinity of these boundaries. In the regime where $\omega>4|\Jeff|$, no resonance can occur at $g=0$, so that the system is stable; upon increasing $g$, a first mode can become unstable, which is the one of maximum $\Eav (q)$, namely $q=\pi$; see Eq.~\eqref{eq:Eav}. Conversely, for $\omega<4|\Jeff|$, there always exists a particular mode fulfilling the resonance condition : the system is unstable at any interaction strength, and the most unstable mode is located at a certain momentum $q<\pi$ (at lowest order, $q^\mathrm{mum}$ is the mode at resonance; see Eq.~\eqref{eq:Eav}). We stress that such conclusions are similar to those found in the non-resonant case~\cite{lellouch2017}.

\subsection{Extension to higher dimensional model}
		\begin{figure}[!t]
			\begin{center}
				\includegraphics[width=8cm]{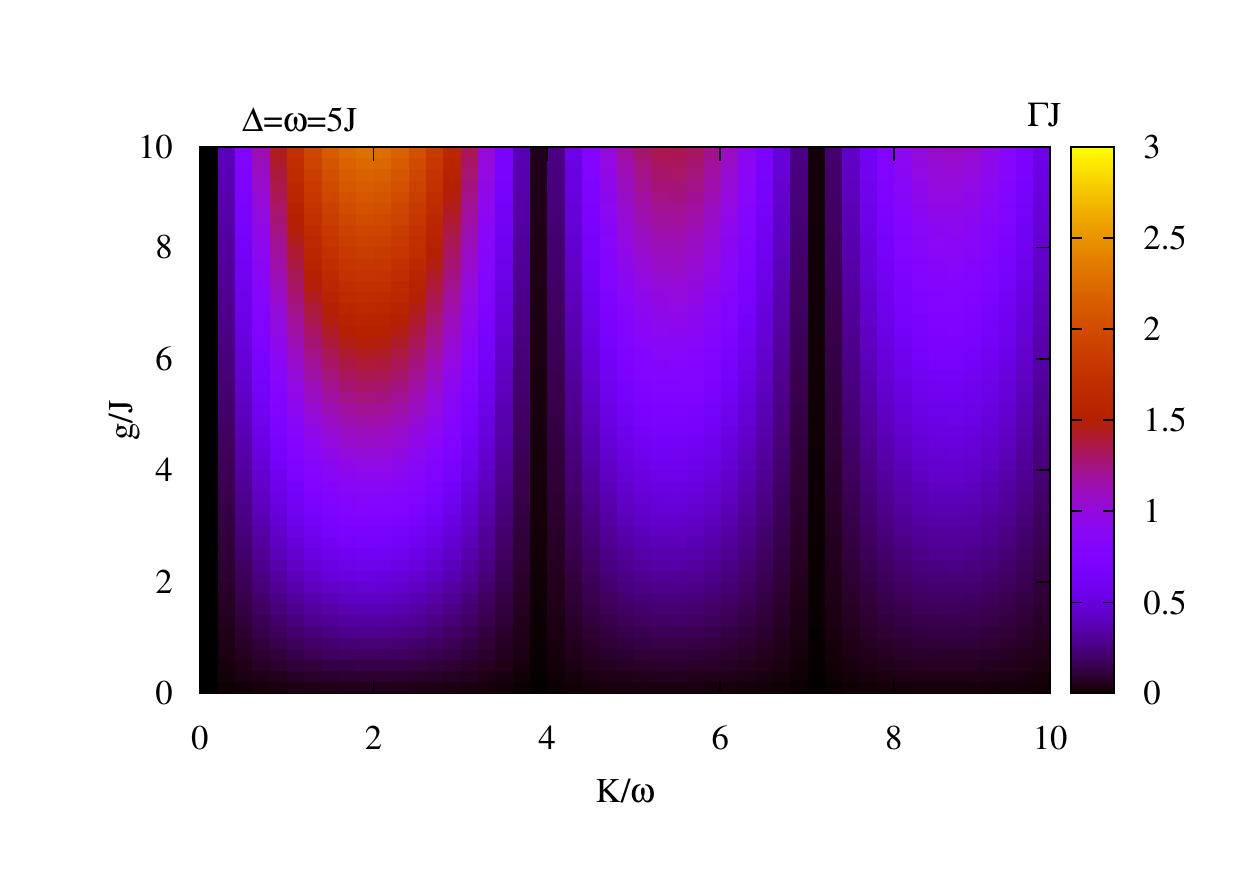}
			\end{center}
			\caption{Stability of the resonantly-driven Wannier-Stark ladder in the presence of a transverse continuous dimension. Shown is the instability rate $\Gamma$ as a function of the interaction strength $g=U\rho$ and the modulation amplitude $K/\omega$. We set $\omega=\Delta=5J$.
				\label{fig:ResCont}
			}
		\end{figure}
The present analysis can be straightforwardly extended to models featuring transverse directions (be it lattice  or continuous directions). 

For instance, in the case where of a continuous transverse degree of freedom is present, we find the stability diagram of Fig.~\ref{fig:ResCont}. As observed in Ref.~\cite{lellouch2017}, the presence of transverse modes enhance the instabilities by opening new instability channels. More precisely, in the case of an unbounded bandwidth, there always exists Bogoliubov modes that are resonant with the drive frequency $\omega$; as a consequence, instabilities occur at any finite interaction strength $g$. We note that, in contrast with the purely 1D case treated above, the cancellation of the effective tunneling does not result in a vanishing of $\Eav$; hence, in this case, the stability regions located near the zeros of $\mathcal{J}_{1}(K/\omega)$ are rather imputable to the cancellation of the perturbation's amplitude $\alpha_q$ that is associated with the underlying (effective) parametric oscillator [Eq.~\ref{eq:appalpha} in Appendix A], and which also scales as $\mathcal{J}_{1}(K/\omega)$.

Finally, in that case, simple analytical formulas are obtained for both $\Gamma$ and $\qbf^\mathrm{mum}$ [see Appendix for details; here the index $x$ denotes the lattice direction]:
	\begin{itemize}
	\item[(i)] If $\omega\!>\!\sqrt{4\vert \Jeff \vert(4\vert\Jeff\vert+2g)}$, one finds
		\begin{equation}
		q_x^\mathrm{mum}=\pi; \quad (q_{\bot}^\mathrm{mum})^2/2m = \sqrt{g^2+\omega^2}-g-4|\Jeff|.
		\label{eq:qmum1}
		\end{equation}
		\begin{equation}
		\Gamma= 2J\left|\mathcal{J}_1(K/\omega)\right|\dfrac{g}{\omega}.
		\label{eq:taux1}
		\end{equation}
		
		\item[(ii)] If $\omega\!<\!\sqrt{4\vert\Jeff\vert(4\vert\Jeff\vert+2g)}$, we find
		\begin{equation}
		q_x^\mathrm{mum}= 2 \arcsin \sqrt{\dfrac{\sqrt{g^2+\omega^2}-g}{4|\Jeff|}} ; \quad q_{\bot}^\mathrm{mum}=0.
		\label{eq:qmum2}
		\end{equation}
		\begin{align}
		\Gamma=(\sqrt{g^2+\omega^2}-g) \dfrac{g}{\omega}.
		\label{eq:taux2}
		\end{align}
	\end{itemize}
We note a strong similarity with the non-resonant case~\cite{lellouch2017}; a notable difference, though, is the fact that the rate $\Gamma$ does not depend on the shaking amplitude $K$ in the low-frequency regime [Eq.~\eqref{eq:taux2}]. These predictions, in particular the existence of two different regimes characterized by very specific dependences of $\Gamma$ and $\qbf^\mathrm{mum}$ on the model parameters, could be readily tested by present-day experiments, providing a clear and unambiguous signature of parametric instabilities in resonantly-modulated systems.

	\section{Moving lattices with a space-dependent phase: The 1D case}
	\label{sec:1Dfluxpi}

\qquad We now consider another type of modulation scheme, which involves a main (primary) optical lattice that is perturbed by a (secondary) moving-lattice potential~\cite{stamper2001spinor,goldman2015a}. Introducing a space-dependent phase in this moving-lattice potential has been shown to generate non-trivial effective gauge fields and topological band structures in the context of 2D neutral gases~\cite{aidelsburger2011,aidelsburger2013,miyake2013,kennedy2015,aidelsburger_14,tai2017microscopy}. Before addressing the case of 2D systems (Section~\ref{sec:2DHof}), where such gauge structures appear, we first investigate the properties of a simpler 1D toy model.

\subsection*{The model}

We consider a 1D model described by the Hamiltonian
 \begin{eqnarray}
 \hat{H} & =  \sum_{n} \Big[ & -J (\ahd_{n+1}\ah_{n}+\text{h.c.}) + \dfrac{U}{2} \ahd_{n}\ahd_{n}\ah_{n}\ah_{n}  + \Delta n \ahd_{n}\ah_{n} \nonumber\\
 & & +  K\cos(\omega t + n \theta) \ahd_{n}\ah_{n} \Big], \quad \theta=\pi ,
 \label{eq:Hpi}
 \end{eqnarray}
where $\ah_{n}$ annihilates a particle at lattice site $n$, $J>0$ denotes the tunneling amplitude of nearest-neighbor hopping, $U>0$ is the repulsive on-site interaction strength, and $\Delta>0$ is the energy difference between two consecutive sites. The second line of Eq.~\eqref{eq:Hpi} captures the effects of the secondary (moving-lattice) potential; the latter is characterized by the amplitude $K$, the frequency $\omega$, and a phase difference of $\theta\!=\!\pi$ between two consecutive sites. In the following, we consider the resonant case where $\omega=\Delta$.

Before analyzing the existence of parametric instabilities in this model, we point out that such moving-lattice potentials generically produce momentum kicks, which can be directly revealed in momentum distributions~\cite{goldman2015a}. However, these effects have a zero average over one period of the drive~\cite{goldman2014a}; in particular, such momentum kicks do not influence the parametric instabilities explored in the present work.

It is convenient to perform our analysis in the rotating frame, defined by the unitary transformation $$R(t)=e^{i\sum_n [\alpha\sin(\omega t + n \pi)\ahd_{n}\ah_{n} + \omega t n \ahd_{n}\ah_{n}]},$$ with $\alpha=K/\omega$. 
In this frame, the Hamiltonian reads
 \begin{eqnarray}
 \hat{H} & = & -J \sum_{n}  [e^{i2\alpha(-1)^{n+1}\sin(\omega t)}e^{i\omega t}\ahd_{n+1}\ah_{n}+\text{h.c.}] \nonumber\\
 & & + \dfrac{U}{2} \sum_{n}\ahd_{n}\ahd_{n}\ah_{n}\ah_{n},
 \label{eq:Hrefpi}
 \end{eqnarray}
so that translational invariance (with a periodicity of two lattice sites) is restored.

\begin{figure}[!h]
			\begin{center}
				\includegraphics[width=8cm]{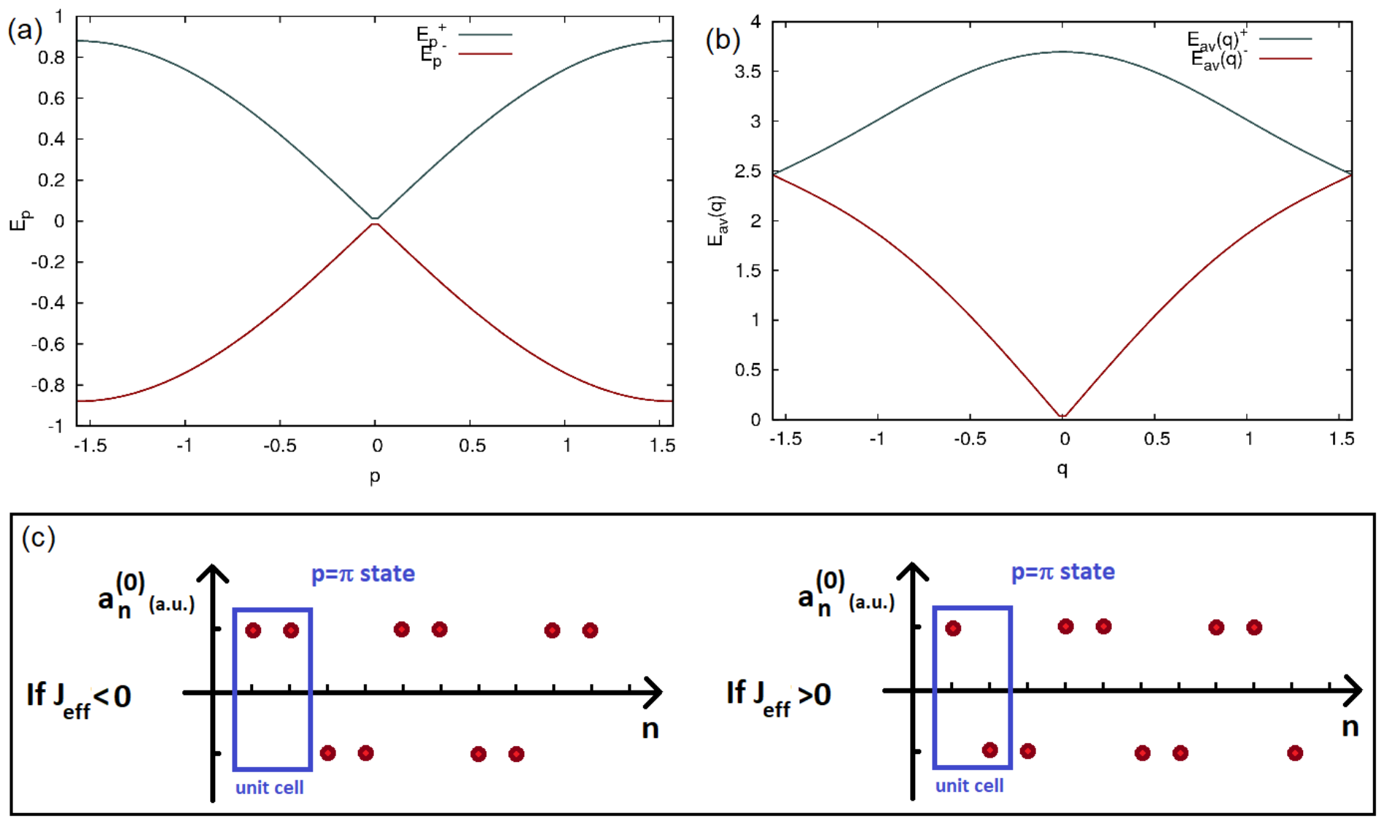}
			\end{center}
			\caption{(a) Single-particle spectrum of the effective Hamiltonian in Eq.~~\eqref{eq:Heffpi} for $J=1$ and $\alpha=0.5$; the ground state corresponds  to the state $p_0=\pi/2$ in the $``-"$ band . (b) Time-averaged Bogoliubov dispersion for the same parameters and $g=3J$. The two branches are reminiscent of those of the single-particle spectrum (the global shift of $\pi/2$ comes from the fact that excitation momenta are defined with respect to the ground-state, which is a $\pi/2$ state), but these are modified by the interactions; in particular, Goldstone modes arise at $q\approx0$ in the lowest band, as generically expected. (c) Qualitative picture of the ground-state in which condensation occur; this state does not depend on the sign of $\Jeff$.
				\label{fig:specHeffpi}
			}
		\end{figure}

In the absence of interactions ($U\!=\!0$), we recall that the Floquet effective Hamiltonian associated with Eq.~(\ref{eq:Hrefpi}) simply reads~\cite{goldman2015a}
 \begin{equation}
 \hat{H}_{\mathrm{eff}}^{(0)}= -\Jeff  \sum_{n}  [(-1)^{n+1}\ahd_{n+1}\ah_{n}+\text{h.c.}] , 
 \label{eq:Heffpi}
 \end{equation}
where $\Jeff=J\mathcal{J}_{-1}(2\alpha)$. The Hamiltonian in Eq.~\eqref{eq:Heffpi} has a periodicity of two lattice sites, so that its spectrum [shown in Fig.~\ref{fig:specHeffpi}(a)] displays two energy bands,
 \begin{equation}
 E_p^{\pm}=\pm 2 |\Jeff \sin p| .
 \label{eq:VepsHeffpi}
 \end{equation}
We note that the corresponding eigenstates are labelled by the quasi-momentum $p$, which is defined in a reduced Brillouin zone $p \in [-\pi/2;\pi/2]$. Here, the ground state corresponds to the state with $p_0\!=\!\pi/2$ in the $``-"$ band; see Fig.~\ref{fig:specHeffpi}(a). Thus, this state has alternate on-site amplitudes from one unit cell to the consecutive one; furthermore, depending on the sign of $\Jeff$, we find that its one-site coefficients are either equal or opposite within each cell; therefore, the ground state is found to be the same, globally, whatever the sign of $\Jeff$ [see Fig.~\ref{fig:specHeffpi}(c)]; this is in striking contrast with the models analyzed in the previous Section~\ref{sec:1DRes} and in Ref.~\cite{lellouch2017}.

\subsection*{Numerical results}

As previously in Section~\ref{sec:1DRes}, the initial state is again taken to be the ground state of the naive ``interacting effective Hamiltonian" [i.e.~Eq.~(\ref{eq:Heff2}) with Eq.~(\ref{eq:Heffpi})], which is obtained numerically through imaginary-time propagation. Interestingly, we find that this state is still characterized by the quasi-momentum $p_0\!=\!\pi/2$, even in the presence of interactions.   \\

\begin{figure}[!h]
			\begin{center}
				\includegraphics[width=8cm]{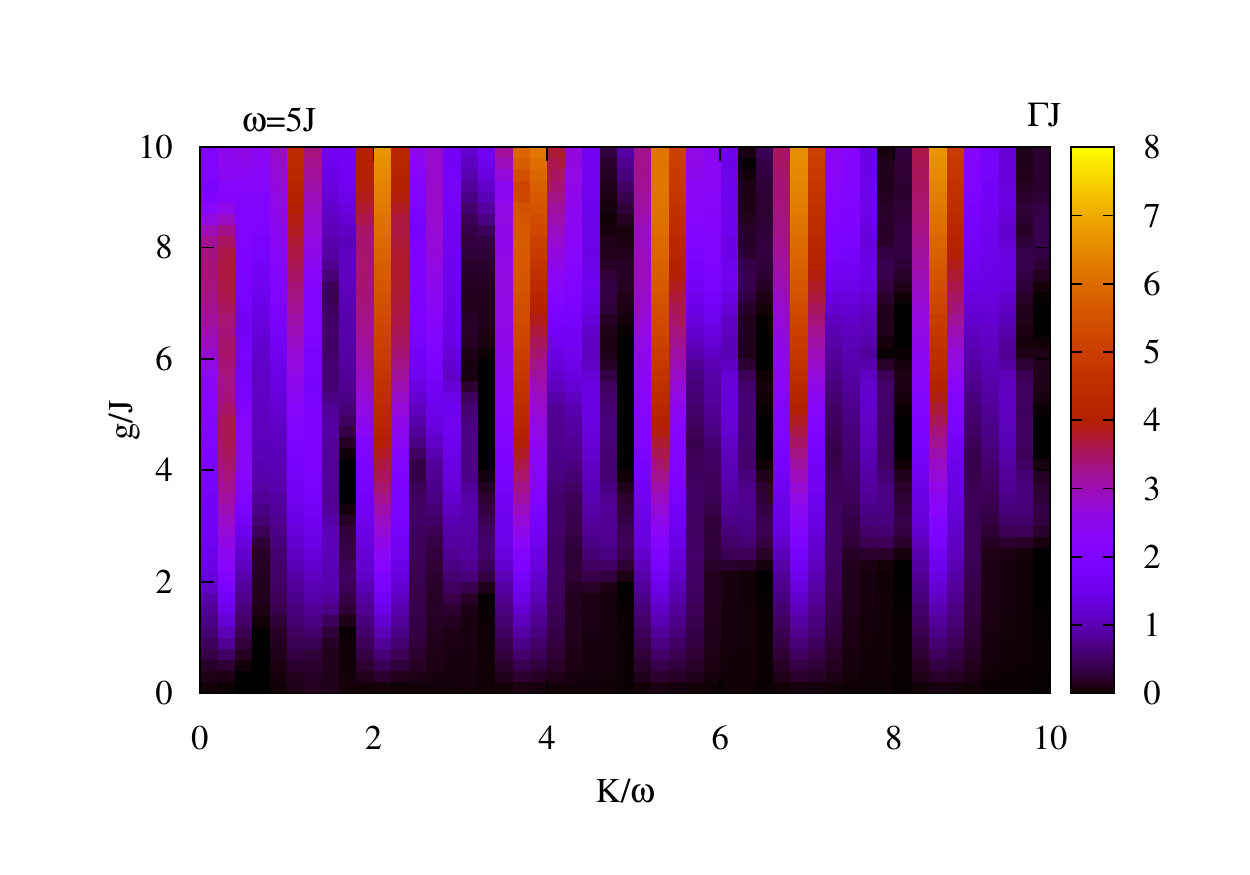}
			\end{center}
			\caption{Stability in the presence of a 1D moving lattice with a spatially-dependent phase. Shown is the numerical instability rate of the model described by Eq.~(\ref{eq:Hpi}) as a function of the interaction strength $g=U\rho$ and the modulation amplitude $K/\omega$; here we set $\omega\!=\!\Delta\!=\!5J$. The quasi-repetition of patterns, as a function of $\alpha=K/\omega$, follows the dependence of the effective tunneling, $J\mathcal{J}_{-1}(2K/\omega)$.
				\label{fig:Num1Dfluxpi}
			}
		\end{figure}

The stability diagram obtained from this initial state is displayed in Fig.~\ref{fig:Num1Dfluxpi}, which shows the instability rate $\Gamma$ as a function of the modulation amplitude and the interaction strength.

The stability diagram in Fig.~\ref{fig:Num1Dfluxpi} is dominated by a quasi ``periodic" structure, as a function of $\alpha=K/\omega$, which arises from the dependence of the effective tunneling along the lattice direction, $J\mathcal{J}_{-1}(2\alpha)$: similarly to what was observed for other modulated band models (see Refs.~\cite{creffield2009,lellouch2017} and Section~\ref{section_analytical}), we note that stable regions are indeed privileged when the effective tunneling $\Jeff\!\approx\!0$, which is consistent with the parametric resonance criterion $\Eav(q)\!=\!\omega$ introduced in Section~\ref{section_analytical}. 

This important observation allows one to anticipate the presence of stable regions in other time-modulation schemes. An immediate extension is the case where the phase difference $\theta$ between consecutive sites [which we took equal to $\theta\!=\!\pi$ in Eq.~\eqref{eq:Hpi}] takes another value. Effective Hamiltonians for such models can be simply obtained using the formulas of Ref.~\cite{goldman2014a}; for instance, considering a moving lattice with a phase difference of $\theta\!=\!\pi/2$, we find an effective tunneling $\Jeff\!=\!J\mathcal{J}_{-1}(4\alpha)$ [see~\cite{goldman2014a} for details], which implies a stability diagram associated with narrower instability lobes. More generally, for models of the form given in Eq.~(\ref{eq:Hpi}) with arbitrary phase $\theta n$, stable zones are obtained whenever $J\mathcal{J}_{-l}(p\alpha)\!\approx\!0$, where $l$ is the order of the resonance ($\Delta\!=\!l\omega$) and where $p$ is the spatial periodicity of the phase $n \theta$ entering the time-modulation (i.e.~the moving lattice).

Besides, our numerical calculations reveal that the onset of instability (i.e.~the boundaries of the stability diagram in Fig.~\ref{fig:Num1Dfluxpi}) is dominated by a most unstable mode, which in this case corresponds to the $q\!=\!0$ mode; we remind that this momentum value is evaluated with respect to the ground-state, as always tacitly implied within our formalism. To understand this, let us compute the time-averaged Bogoliubov dispersion $\Eav(q)$ for the present model, which yields 
\begin{widetext}
 \begin{eqnarray}
\Eav(q)= \sqrt{4|\Jeff|^2(1+\cos^2 q)+4|\Jeff|g \pm \sqrt{16|\Jeff|^2 g^2 \cos^4 q + 64|\Jeff|^3\cos^2 q(|\Jeff|+g)}}.
 \label{eq:Eavpi}
  \end{eqnarray}
 \end{widetext}
As shown in Fig.~\ref{fig:specHeffpi}(b), this dispersion is made of two branches, which are reminiscent of the two-branch single particle spectrum~\footnote{The global shift of $\pi/2$ comes from the fact that excitation momenta are defined with respect to the ground-state, which is a $\pi/2$ state in the present case.}, but are modified by the interactions. In particular, Goldstone modes arise at $q\approx0$ in the lowest band, as generically expected as a result from the broken $U(1)$ symmetry~\cite{goldstone1962}. The absolute maximum is reached for $q\!=\!0$ in the upper branch, which indeed precisely corresponds to the most unstable mode identified in our numerics:~similarly to the previous model discussed in Section~\ref{sec:1DRes}, we thus find that the onset of parametric instability is governed by the mode of highest time-averaged Bogoliubov energy, which is consistent with the fact that this mode will be the first one to resonate [$\Eav(q\!=\!0)\!=\!\omega$] as one increases the interaction strength $g$~\cite{lellouch2017}. We emphasize that this most unstable mode differs from the one identified for the resonantly-shaken Wannier-Stark ladder of Section~\ref{sec:1DRes}. \\
		
It is remarkable that the conclusions emanating from our analysis of the moving lattice are qualitatively similar to those related to other shaken-lattice models; see Sec.~\ref{sec:1DRes} and Ref.~\cite{lellouch2017}. This suggests a reliable and intuitive guideline to predict stable zones of the stability diagram and to identify the most unstable mode, based on the simple knowledge of the effective band structure (i.e.~the effective tunneling and the time-averaged Bogoliubov dispersion). 

\section{The driven-induced Harper-Hofstadter model}
	\label{sec:2DHof}

In this Section, we extend our previous analysis [Section~\ref{sec:1Dfluxpi}] to a two-dimensional setting, which has been exploited to realize the Harper-Hofstadter Hamiltonian in cold atoms~\cite{aidelsburger2011,aidelsburger2013,miyake2013,kennedy2015,aidelsburger_14,tai2017microscopy}.

	\subsection*{The model}
		\begin{figure}
			\begin{center}
				\includegraphics[width=8.3cm]{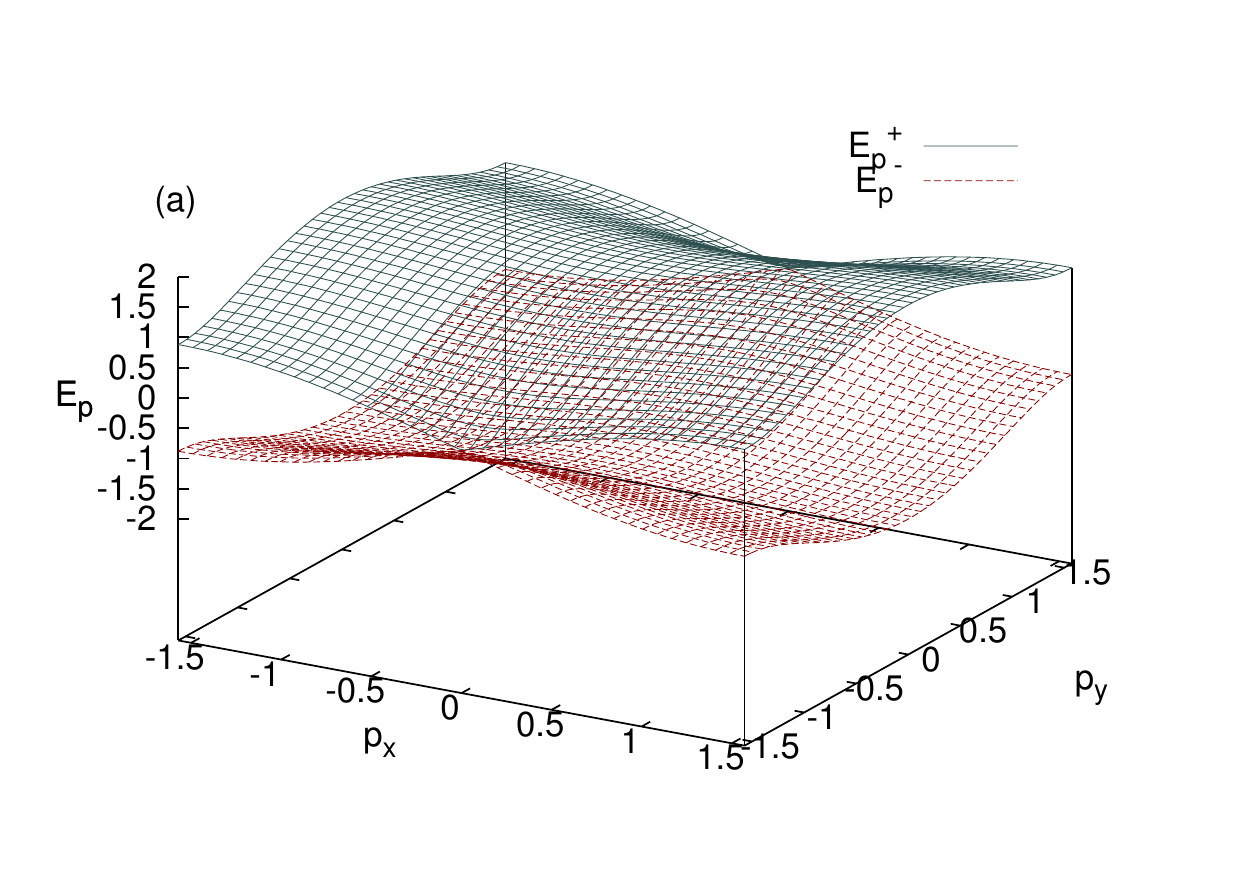}
				\includegraphics[width=8.3cm]{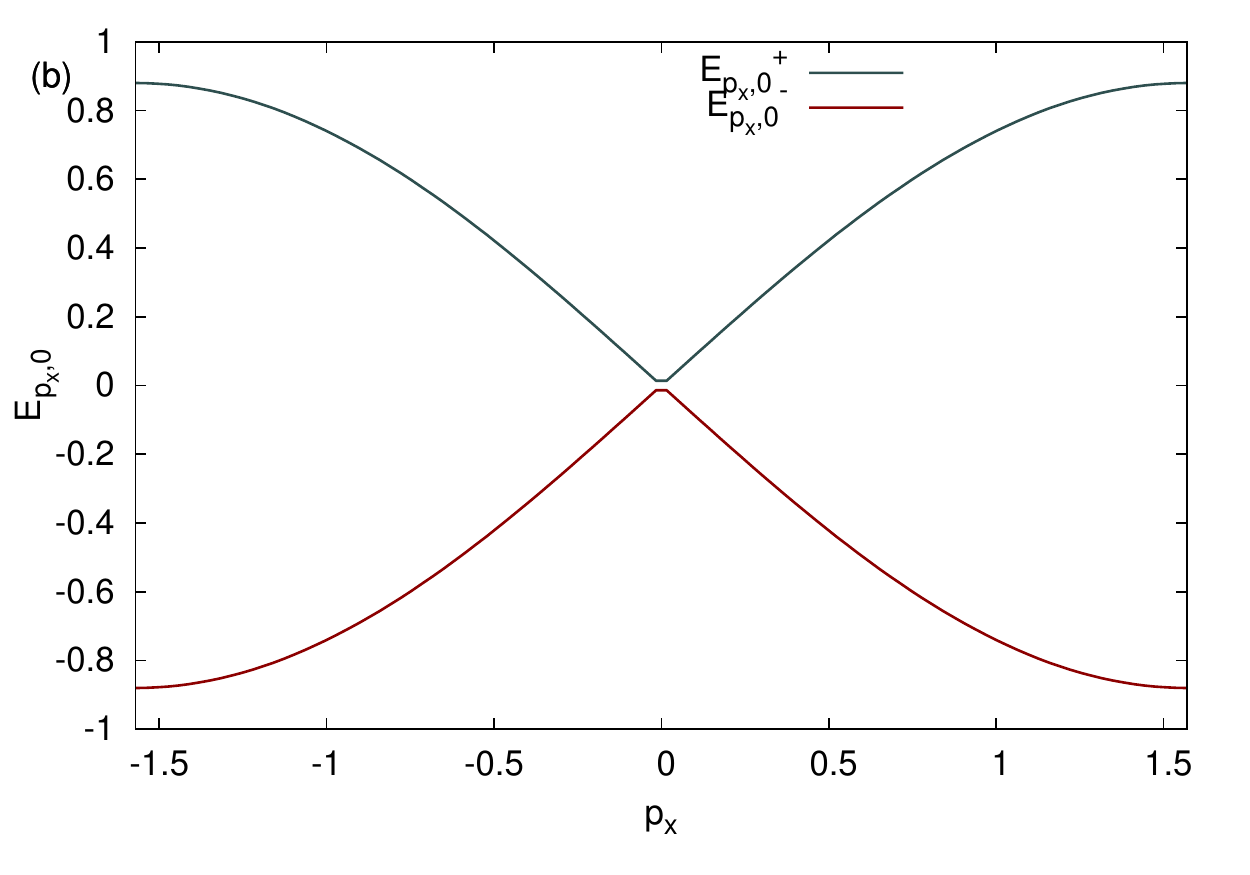}
				\includegraphics[width=8.3cm]{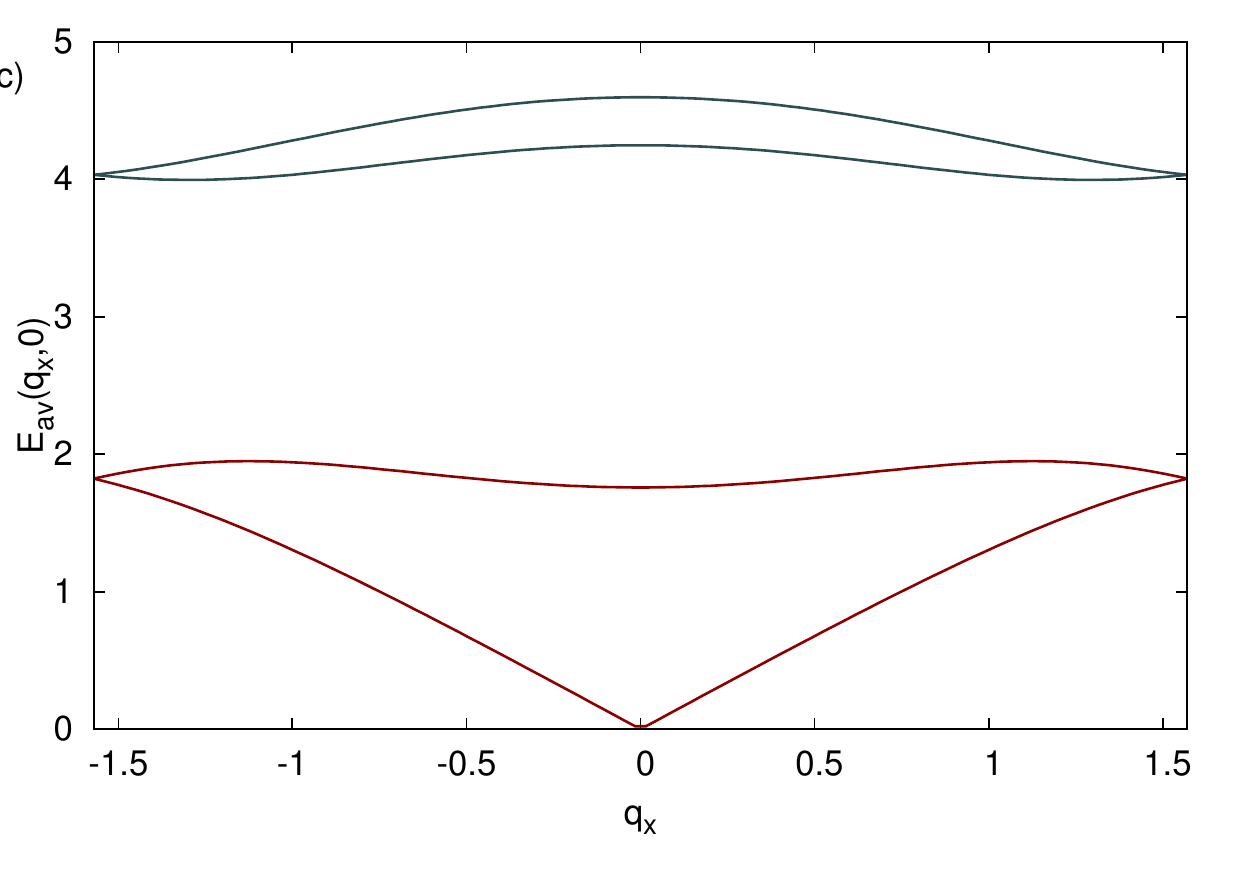}
			\end{center}
			\caption{(a) Single-particle spectrum of the effective Hamiltonian in Eq.~\eqref{eq:Heff2D} for $J_{x,y}=1$ and $\alpha=0.5$; each branch is twofold degenerate, and the two ground-states correspond to states at $p_x=\pi/2, p_y=0$ in the $``-"$ band; (b) Cut of the single-particle spectrum at $p_y=0$; (c) Time-averaged Bogoliubov dispersion for the same parameters and $g=3J$, plotted as a function of $q_x$ at fixed $q_y=0$. The four branches are reminiscent of the two branches of the single-particle spectrum, with a lifting of their twofold degeneracy through interactions; the global shift of $\pi/2$ in the $x$-direction comes from the fact that excitation momenta are defined with respect to the ground-state, which is a $p_x=\pi/2$ state; in particular, Goldstone modes arise at $q\approx0$ in the lowest band, as generically expected.
				\label{fig:specHeff2D}
			}
		\end{figure}
We consider a two-dimensional extension of the previous model [Eq.~\eqref{eq:Hpi}], which we define by the Hamiltonian
 \begin{eqnarray}
 \hat{H} & = \sum \limits_{m,n} \Big\{ & -J_x (\ahd_{m+1,n}\ah_{m,n}+\text{h.c.}) -J_y (\ahd_{m,n+1}\ah_{m,n}+\text{h.c.})\nonumber\\
 & & + \big [ K\cos(\omega t + m \theta_x +n \theta_y) + \Delta m \big ]\ahd_{m,n}\ah_{m,n} \nonumber\\ 
 & & + \dfrac{U}{2} \ahd_{m,n}\ahd_{m,n}\ah_{m,n}\ah_{m,n} \Big\}, \quad \theta_x=\theta_y=\pi,
 \label{eq:H2D}
  \end{eqnarray}
where $\ah_{m,n}$ annihilates a particle at lattice site $(m,n)$, $J_x$ (resp.~$J_y$) is the tunneling along the $x$ (resp.~$y$) direction, and $U$ models on-site repulsive interactions. A linear (Wannier-Stark) potential is introduced along the $x$-direction; besides the model features a moving lattice of amplitude $K$, frequency $\omega$ and a $\pi$-phase dependence along both directions ($\theta_{x,y}\!=\!\pi$). In the following, we set the resonance condition $\omega=\Delta$. This model, and other variants, have been realized in several recent ultracold-atom experiments~\cite{aidelsburger2011,aidelsburger2013,miyake2013,kennedy2015,aidelsburger_14,tai2017microscopy}. Here, we focus on the ``$\pi$-flux" configuration~\cite{miyake2013,kennedy2015}, which corresponds to setting $\theta_y\!=\!\pi$ in Eq.~\eqref{eq:H2D}, but we note that several of our results hold for other values of the synthetic flux (see below).\\

In the absence of interactions ($U\!=\!0$), the effective Hamiltonian associated with Eq.~\eqref{eq:H2D} reads~\cite{goldman2014a}
 \begin{eqnarray}
  \hat{H}_{\mathrm{eff}} & =  & - \Jeff^x \sum_{m,n} [ (-1)^{m+n+1}\ahd_{m+1,n}\ah_{m,n}+\mathrm{h.c.}] \nonumber\\
  & & - \Jeff^y \sum_{m,n}[\ahd_{m,n+1}\ah_{m,n}+\text{h.c.}],
 \label{eq:Heff2D}
  \end{eqnarray}
with $\alpha=K/\omega$, $\Jeff^x=J_x\mathcal{J}_{-1}(2\alpha)$, and $\Jeff^y=J_y \mathcal{J}_{0}(2\alpha)$. This corresponds to the Harper-Hofstadter model~\cite{hofstadter1976}, with a magnetic flux $\Phi\!=\!\pi$ in each unit cell, and with different effective tunnelings along the $x$ and $y$ directions. As stated above, other choices for the moving-lattice phase ($\theta_y\!\ne\!\pi$) lead to different fluxes per unit cell~\cite{aidelsburger2013,aidelsburger_14,goldman2014a}. We recall that the Harper-Hofstadter model displays the well-known ``Hofstadter's butterfly" spectrum~\cite{hofstadter1976}, a rich fractal structure that hosts Dirac semimetals and Chern insulating phases~\cite{kohmoto1989zero}. In this sense, at the single-particle level, the time-dependent model under consideration [i.e.~Eq.~\eqref{eq:H2D} with $U\!=\!0$] is one of the simplest systems realizing artificial gauge fields and topological band structures for neutral atoms in 2D optical lattices~\cite{aidelsburger2011,aidelsburger2013,miyake2013,kennedy2015,aidelsburger_14,tai2017microscopy}; we point out that a BEC was recently realized in the ``$\pi$-flux" configuration of this model, which further motivates the present study~\cite{kennedy2015}.

\subsection*{Numerical results}

	In order to extract the instability properties of the 2D model in Eq.~\eqref{eq:H2D}, we now apply the same procedure as for the 1D model of Section~\ref{sec:1Dfluxpi}. The initial state is again chosen as the ground state of the naive ``interacting effective Hamiltonian"; however, we point out that a complexity arises here from the degeneracy of this condensation state~\cite{hofstadter1976,powell2011}.
		
To see this, one observes that the single-particle effective Hamiltonian in Eq.~(\ref{eq:Heff2D}) is periodic over a $2 \times 2$ cell, so that its eigenstates are labelled by the quasi-momentum $\boldsymbol{p}$ defined in the reduced Brillouin zone $p_x,p_y \in [-\pi/2;\pi/2]$. The spectrum features two energy bands (labelled $\pm$)
 \begin{equation}
 E_{\mathbf{p}}^{\pm}=\pm 2 \sqrt{{\Jeff^x}^2 \sin^2 p_x + {\Jeff^y}^2 \cos^2 p_y},
 \label{eq:VepsHeff2D}
 \end{equation}
which are both twofold degenerate [see Figs.~\ref{fig:specHeff2D}(a-b)]. The ground state corresponds  to the state with $\boldsymbol{p}\!=\!(\pi/2, 0)$ in the $``-"$ band, and is twofold degenerate~\footnote{Let us stress that the effective Hamiltonian in Eq.~(\ref{eq:Heff2D}) obtained from Eq.~(\ref{eq:H2D}), does not correspond to the Landau gauge, hence the difference compared to the more common version of the Harper-Hofstadter model whose spectrum is  $E_{\mathbf{p}}^{\pm}=\pm 2 \sqrt{{\Jeff^x}^2 \cos^2 p_x + {\Jeff^y}^2 \cos^2 p_y}$ and ground-state in $p_x=p_y=0$.}. 
 
 In the presence of interactions, the ground state still features this twofold degeneracy~\cite{powell2011}, as well as the momentum characteristic $(p_x=\pi/2, p_y=0)$. Therefore, several choices can \textit{a priori} be made for the initial state of our analysis, within the whole degenerate ground space, as we now explore.

\subsubsection*{Stability diagram and sensitivity to the ground state}

\begin{figure}[!t]
	
	\begin{minipage}{8cm}
		\includegraphics[width=8.3cm]{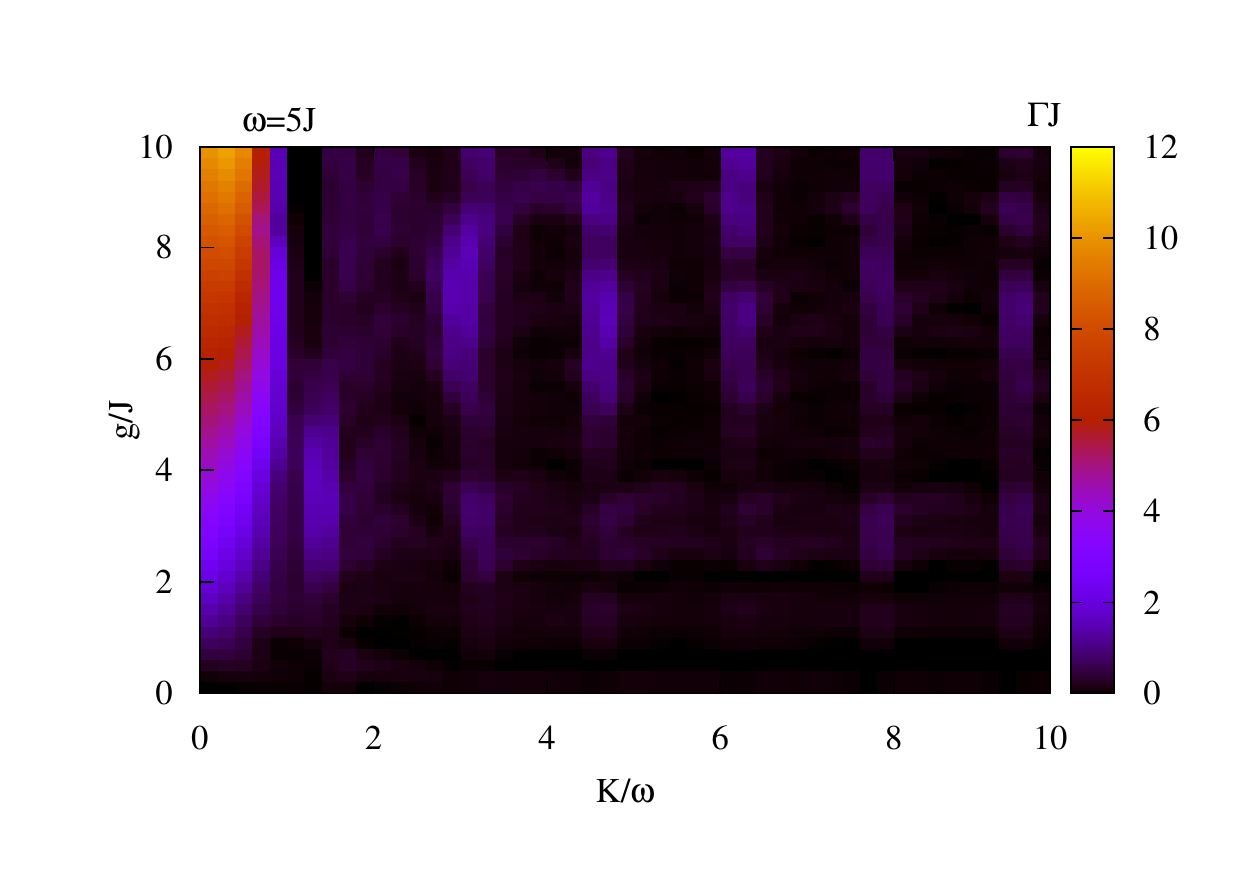}
	\end{minipage}
	\begin{minipage}{8cm}
		\centering
		\includegraphics[width=8.3cm]{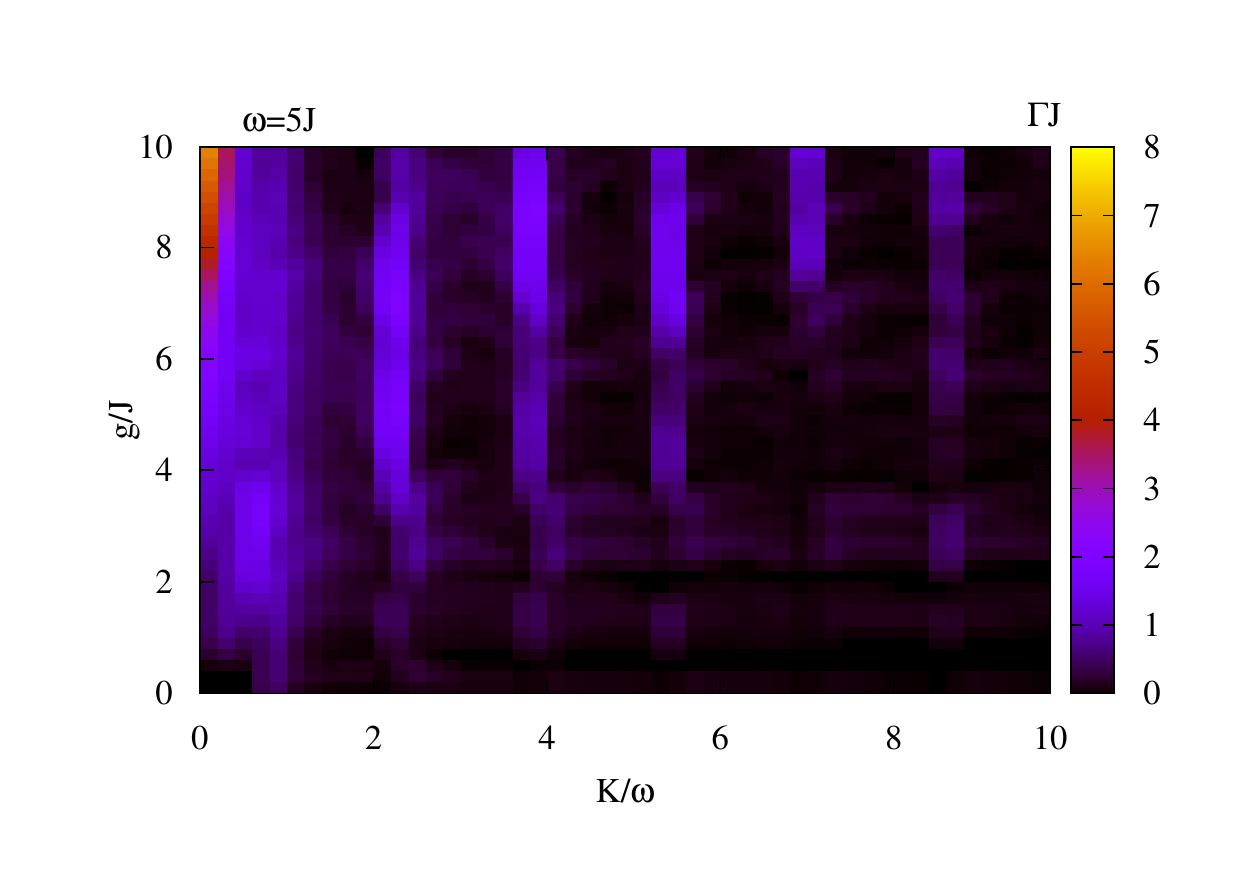}
	\end{minipage}
\vspace{0.5cm}
	\caption{Stability diagram for the driven-induced Harper-Hofstadter model described by Eq.~(\ref{eq:H2D}), as a function of the modulation amplitude $K/\omega$ and the interaction strength $g$; we set frequency $\omega\!=\!5J$ and $J_{x,y}\!=\!J$. The two panels correspond to the stability diagrams obtained for two orthogonal ground-states.
      \label{fig:num2DHof}}
\end{figure}
Figure~\ref{fig:num2DHof} shows the instability diagrams obtained for two different orthogonal ground-states in this subspace. At first sight, they are found to be very similar in the sense that the rates are of the same order of magnitude, and the diagrams feature very similar structures as a function of the model parameters; in particular, the same ``periodicity" as in Fig.~\ref{fig:Num1Dfluxpi} is observed as a function of modulation amplitude $\alpha=K/\omega$. Yet, the exact positions of the stable and unstable regions are shifted, reflecting that the later are very sensitive to the ground-state in which the system is prepared. 

The later observation that stable regions cannot be simply obtained by analyzing the behavior of the effective tunneling $J_{\text{eff}}^{x,y}$ [Section~\ref{section_analytical}] can be related to the fact that these quantities never cancel simultaneously; indeed, in this specific 2D model, the time-averaged Bogoliubov dispersion never vanishes, meaning that the simple ``flat-band" criterion of Section~\ref{section_analytical} cannot be applied (a priori, there might always exist some modes fulfilling the resonance condition $\Eav=\omega$). Consequently, we conjecture that the stable zones in Fig.~\ref{fig:Num1Dfluxpi} should rather be governed by the amplitude of the perturbation entering the underlying effective parametric-oscillator model [i.e. the analogue of the quantity $\alpha_{\bf{q}}$ that was defined in Eq.~\ref{eq:appalpha}, in Appendix A, for the model of Sec.~\ref{sec:1DRes}]; although its full analytical derivation cannot be obtained in the present case, it is natural to believe that this amplitude is indeed ground-state dependent, since the Bogoliubov treatment leading to this effective parametric-oscillator model relies on the actual condensation state that is chosen within the degenerate ground manifold.

\subsubsection*{Most unstable mode}

In contrast, we find that the most unstable mode associated with the onset of the instability is robust with respect to the ground-state. Indeed, independently of the ground state, and within the whole parameter range of the diagram, we find that this  most unstable mode always corresponds to the $\qbf\!=\!0$ mode (with respect to the ground-state). This can again be accounted for by the fact that it is the mode of highest time-averaged Bogoliubov energy. To verify this, we calculate the time-averaged Bogoliubov spectrum $\Eav$, which is obtained by numerically diagonalizing the Bogoliubov Hamiltonian derived from the GPE combined with the static effective Hamiltonian in Eq.~\eqref{eq:Heff2D}; we note that analytics exist in the symmetric case where $\Jeff^x=\Jeff^y$; see~\cite{powell2011}. The resulting Bogoliubov spectrum $\Eav$ is plotted in Fig.~\ref{fig:specHeff2D}(c). Consistently with the general features reported in~\cite{powell2011}, this spectrum is made of four branches (the degeneracy observed at the single-particle level being lifted) and indeed has its absolute maximum at $\qbf\!=\!0$ in the upper branch.

Interestingly, this behavior is in fact expected to be very generic. For a model configuration leading to a flux $\Phi\!=\!1/3$, i.e.~$\theta_y\!=\!\pi/3$, we also find a most unstable mode at $\qbf=0$, consistently with the fact that it is the mode of highest time-averaged Bogoliubov energy [see Ref.~\cite{powell2011} for the Bogoliubov spectrum in that case].\\

We conclude this Section by noting that the stability diagram of the driven-induced Harper-Hofstadter model [Fig.~\ref{fig:num2DHof}] displays relatively large stability regions, which extend up to significant values of the  interaction strength $g$, hence suggesting potentially favorable regimes of operation (as far as parametric instabilities are concerned). However, one should point out that current experiments typically feature a transverse (``tube") direction, which is generically expected to trigger or increase instabilities~\cite{lellouch2017}. Besides, our result that the stability diagram strongly depends on the prepared ground state appears as an important feature of these time-modulated systems, which should be taken in account in experiments. 

\section{Conclusion}\label{conclusion}
In this work, we analyzed the parametric instabilities that occur in a variety of resonantly-driven band models.
While the general method used to identify and characterize these instabilities had already been applied to non-resonant models~\cite{lellouch2017,creffield2009}, we have hereby demonstrated its usefulness and flexibility to address the relevant case of resonant time-modulations. \textit{Ab initio} predictions for instability rates and for the most unstable mode (which appears to be the most robust and directly accessible signature of parametric instabilities) were obtained; these could be directly tested in current ultracold-atom experiments~\cite{aidelsburger2011,aidelsburger2013,miyake2013,kennedy2015,aidelsburger_14,tai2017microscopy}, for instance, in view of optimizing their operating regimes.

Our study has confirmed the generic role played by parametric resonances in the instabilities that appear in these driven system, and which directly involve the drive frequency and the dispersion of the Floquet-Bogoliubov modes. In particular, while genuine analytical results can only be obtained in simple cases, many qualitative features of these instabilities appear to be generic, hence providing an intuitive picture that could guide experiments to stable regimes of operation. 

On the one hand, instability rates exhibit a dependence on the modulation amplitude that is, in most cases, governed by the effective tunneling, and favorable regions are generally found whenever this effective tunneling is weak (based on the resonance criterion $\Eav\!=\!\omega$ underlying parametric instabilities and the fact that $\Eav\!\approx\!0$ in these regions); while this conclusion is justified in models where all effective tunneling amplitudes can simultaneously vanish (as in the simple 1D models discussed in this work), it should however be treated with care in higher dimensions, where stability regions seem to be rather governed by the amplitude of the perturbation entering the underlying effective parametric-oscillator model [i.e. the analogue of the quantity $\alpha_{\bf{q}}$ in Eq.~\ref{eq:appalpha} in Appendix A]; in particular, in the presence of degenerate ground states, as in the ``Harper-Hofstadter" case treated in Section~\ref{sec:2DHof}, the instability rates were found to be very sensitive to the actual ground-state in which the system was prepared. Besides, we note that the energy scales of the (effective) system mostly rely on the effective tunneling, which indicates that a compromise must be found in actual implementations (in the context of fractional Chern insulators, it is crucial to maximize the size of topological gaps in view of generating strongly-correlated states in the presence of finite temperature). 

On the other hand, for models exhibiting a finite bandwidth, the most unstable mode responsible for the onset of instability is always found to be that of highest effective Bogoliubov energy $\Eav$ (for experimental values of the drive frequency~\cite{lellouch2017}). When a continuum is present and the dispersion is unbounded, the most unstable mode is found (at lowest order) to fulfill the resonance condition $\Eav(\qbf^\mathrm{mum})\!=\!\omega$. Therefore, the simple knowledge of the effective band structure [$\Jeff$ and $\Eav$] allows one to anticipate such behaviors, which are expected to be directly relevant to experiments.


Altogether, this work constitutes a first step in view of controlling and exploiting the potentialities of interacting modulated BECs. In particular, achieving stable BECs in the context of time-modulated optical lattices, with reduced instabilities and losses, will constitute a first step towards the cold-atom realization of strongly-correlated states of matter with topological features, such as fractional Chern insulators~\cite{regnault2011fractional,sterdyniak2013series,bergholtz2013topological,grushin_14,moller2015fractional}. However, many open problems remain. First of all, this study was performed it the mean-field interaction regime, which is expected to give access to the short-term dynamics. Longer-time dynamics, including the possibility for thermalization, are dominated by non-linear couplings between the excitation modes and the condensate, which go beyond the present Bogoliubov treatment. Moreover, the analysis presented in this work neglects the effects of higher bands, which are indeed expected to be weak when setting the drive frequency within a gap of the (effective) spectrum; however, higher-band effects are also expected to become important at longer times. Then, independently from the driving scheme itself and the obtained stability properties, a central question is that of the preparation of the initial state, as loading the system into a desired eigenstate with highest fidelity is far from being trivial: one solution could be to apply adiabatic perturbation theory in the presence of the periodic drive~\cite{pweinberg_15,novicenko2016}, but there might exist other alternatives and their impact on the stability properties of the prepared state remains uncharacterized. Finally, the interplay between parametric instabilities and other instability mechanisms neglected in our approach, especially inter-band transitions~\cite{weinberg2015,strater2016,reitter2017}, is expected to lead to rich behaviors that still remain to be studied.

	\section*{Acknowledgments}
	We thank M. Aidelsburger, M. Bukov and M. Cheneau for insightful discussions.
	This research was financially supported by the Belgian Science Policy Office under the Interuniversity Attraction Pole project P7/18 ``DYGEST", the FRS-FNRS (Belgium) and the ERC TopoCold Starting Grant. 
	\\


%

	
\newpage

\onecolumngrid

\appendix	

		\section{Analytical treatment of the resonantly-driven Wannier-Stark ladder}
		\label{sec:app1}
		
We present the analytical derivation of instability rates for the resonantly-driven Wannier-Stark ladder of Sec.~\ref{sec:1DRes}. 
For the sake of generalness, we consider an even more general situation where the modulation has a non-trivial uniform phase, and with possible additional transverse directions in the model.
We thus deal with the Hamiltonian
 \begin{equation}
  \hat{H}(t) =-J\sum_{n,\rp} (\ahd_{n+1,\rp}\ah_{n,\rp}+ \mathrm{h.c.})+ \hat{H}_\perp + \frac{U}{2} \sum_{n,\rp} \ahd_{n,\rp}\ahd_{n,\rp}\ah_{n,\rp}\ah_{n,\rp} + \sum_{n,\rp} \Big[\Delta n \ahd_{n,\rp}\ah_{n,\rp} + K\cos(\omega t+\phi) n \ahd_{n,\rp}\ah_{n,\rp} \Big],
 \label{eq:appH}
 \end{equation}
where $\ah_{n,\rp}$ annihilates a particle at lattice site $n$ and transverse position $\rp$, $J>0$ denotes the tunneling amplitude of nearest-neighbor hopping along the x-direction, $\hat{H}_\perp$ describes a kinetic part along transverse directions (be it a lattice or a continuous one), and $U>0$ is the repulsive on-site interaction strength. The on-site potential term is made of a Wannier-Stark ladder along the x-direction introducing an energy shift $\Delta>0$ between consecutive sites, and a time-periodic modulation along the x-direction of amplitude $K$, phase $\phi$, and frequency $\omega=2\pi/T$. The frequency modulation $\omega$ is chosen to be resonant with the offset $\Delta$ ($\Delta=l\omega$) with $l$ denoting some integer.\\

We will perform all the theoretical analysis in the rotating frame (where translational invariance is restored), which is defined by the unitary transformation $\hat{R}(t)\!=\!\e^{-i[\alpha\sin(\omega t+\phi)+\Delta t]\hat{X}}$ (with $\hat{X}\equiv\Sigma_n n \ahd_n\ah_n$ the position operator on the lattice).

\subsection{Condensate dynamics and Bogoliubov equations}

The single-particle quasienergy spectrum of this model reads
 \begin{equation}
  \epsilon_{\qbf} = -2J \Jeff \cos(q_x+l\phi) + \epsilon^\perp(\qp)
 \label{eq:appQen}
 \end{equation}
where $\Jeff=\mathcal{J}_{-l}(\alpha)$ is the effective tunneling (with $\alpha=K/\omega$) and $\epsilon^\perp$ is the dispersion associated with transverse directions.
The effective ground-state, which is taken as the initial state in which the system is supposed to be prepared, is a Bloch state $e^{i(p_xn+\pp\rp)}$ of momentum $\pp=0$, and $p_x\!=\! -l\phi$ if $\Jeff\!>\!0$ or $p_x\!=\!\pi-l\phi$ if $\Jeff\!<\!0$. 

Taking this state, normalized to the condensate density $\rho$, as our initial condition $a_{n,\rp}^{(0)}(t=0)$ for the propagation of the time-dependent Gross-Pitaevskii equation (tGPE), we find that the solution of the (tGPE) reads
 \begin{equation}
 a_{n,\rp}^{(0)}(t)=\sqrt{\rho}e^{ip_xn}e^{-i\Delta t n-i\alpha\sin(\omega t +\phi)n}e^{2iJ[\cos p_x S(t)-\sin p_x C(t)]}e^{-\epsilon^\perp(0)}e^{-i g t}
 \label{eq:appGSt}
 \end{equation}
where we have introduced the functions 
$$S(t)=\sum_{m=-\infty}^{\infty}\dfrac{\sin[(m\omega+\Delta)t+m\phi]-\sin(m\phi)}{m\omega+\Delta}\mathcal{J}_{m}(\alpha)$$ and
$$C(t)=\sum_{m=-\infty}^{\infty}\dfrac{\cos[(m\omega+\Delta)t+m\phi]-\cos(m\phi)}{m\omega+\Delta}\mathcal{J}_{m}(\alpha).$$
Plugging it into the Bogoliubov-De Gennes equations, and taking advantage of the translational invariance to rewrite these in momentum space, we find that the dynamics of a momentum mode $q$ is governed by the equation
	\begin{equation}
		i \partial_t \left( \begin{matrix} u_q  \\ v_q \end{matrix} \right)=\left( \begin{matrix}  \varepsilon(\mathbf{q},t)+g & g        \\ -g & -\varepsilon(-\mathbf{q},t)-g \end{matrix} \right)\left( \begin{matrix} u_q  \\ v_q \end{matrix} \right),
		\label{eq:appBdGE}
	\end{equation}
	where $$\varepsilon(\mathbf{q},t)=4J\sin\dfrac{q_x}{2}\sin\left(\dfrac{q_x}{2}+p_x-\alpha\sin(\omega t+\phi)-\Delta t \right) + \epsilon^\perp_{\qp}.$$ 

\subsection{Mapping on a parametric oscillator model}

Proceeding as in~\cite{lellouch2017}, we then introduce the two following changes of basis: first,
		\begin{equation}
			\left( \begin{matrix} u_\qbf  \\ v_\qbf \end{matrix} \right)=\left( \begin{matrix} \mathrm{cosh}(\theta_\qbf) & \mathrm{sinh}(\theta_\qbf)        \\ \mathrm{sinh}(\theta_\qbf) & \mathrm{cosh}(\theta_\qbf) \end{matrix} \right)\left( \begin{matrix} u'_\qbf  \\ v'_\qbf \end{matrix} \right),
			\label{eq:appchange1}
		\end{equation}
		where $\mathrm{cosh}(2\theta_\qbf)\equiv (\epsav(\qbf)+g)/\Eav(\qbf)$ and $\mathrm{sinh}(2\theta_\qbf)\equiv g/\Eav(\qbf)$ with 
		\begin{equation}
		\Eav(\qbf)=\sqrt{\epsav(\qbf)(\epsav(\qbf)+2g)}
			\label{eq:appEav}
		\end{equation}
	the time-averaged Bogoliubov dispersion and 
	$$\epsav(\qbf)=4|\Jeff|\sin^2(q_x/2)+ \epsilon^\perp_{\qp};$$ and then,
		\begin{equation}
			\left( \begin{matrix} \tilde{u}'_\qbf  \\ \tilde{v}'_\qbf \end{matrix} \right)=\left( \begin{matrix} e^{2i\Eav(\qbf)t} & 0       \\ 0 & 1 \end{matrix} \right)\left( \begin{matrix} u'_\qbf  \\ v'_\qbf \end{matrix} \right).
			\label{eq:appchange}
		\end{equation}
		This allows us to write the Bogoliubov equations under the form
		\begin{equation}
			i\partial_t \left( \begin{matrix} \tilde{u}'_\qbf  \\ \tilde{v}'_\qbf \end{matrix} \right)=\left[ \Eav(\qbf)\hat{\mathbf{1}}+\hat{W}_\qbf(t)+\mathrm{sinh}(2\theta_\qbf)\left( \begin{matrix} 0 & h_\qbf(t)e^{-2i\Eav(\qbf)t}        \\ -h_\qbf(t)e^{2i\Eav(\qbf)t} & 0 \end{matrix} \right)\right] \left( \begin{matrix} \tilde{u}'_\qbf  \\ \tilde{v}'_\qbf \end{matrix} \right).
			\label{eq:appBdGEPO}
		\end{equation}
where $\Eav(\qbf)$ is the time-averaged Bogoliubov dispersion associated with the effective GPE, within the Bogoliubov approximation [see Eq.~(\ref{eq:appEav})], $\hat W_\qbf(t)$ is a diagonal matrix of zero average over one driving period,
		\begin{equation}
			\hat W_\qbf(t)=4J\mathrm{sin}\dfrac{q_x}{2}\sum_{m\neq 0}\mathcal{J}_m(\alpha)\left( \begin{matrix} \mathrm{cosh}2\theta_\qbf\mathrm{sin}\dfrac{q_x}{2}\cos[(m+l)(\omega t+\phi)]-\mathrm{cos}\dfrac{q_x}{2}\sin[(m+l)(\omega t+\phi)]       \qquad \qquad \qquad 0 \qquad \qquad \qquad \\ \qquad \qquad \qquad 0 \qquad \qquad \qquad   \mathrm{cosh}2\theta_\qbf\mathrm{cos}\dfrac{q_x}{2}\sin[(m+l)(\omega t+\phi)]+\mathrm{sin}\dfrac{q_x}{2}\cos[(m+l)(\omega t+\phi)] \end{matrix} \right),\notag
			\label{eq:appW}
		\end{equation}
 and $h_\qbf(t)$ is a (real-valued) function, 
	\begin{align}
		h_\qbf(t)&=4J\sin^2(q_x/2)\sum_{m\neq 0} \mathcal{J}_m(\alpha)\cos[(m+l)(\omega t+\phi)].
		\label{eq:apphq}
	\end{align}
	
As shown in~\cite{lellouch2017}, Eqs.~\ref{eq:appBdGEPO} are formally equivalent to a set of independent parametric oscillators (one for each mode $\qbf$), since they describe harmonic oscillators of eigenfrequencies $\Eav(\qbf)$ driven by a weak time-periodic modulation of frequency $\omega$. As can be anticipated from a RWA treatment of  Eq.~\ref{eq:appBdGEPO}, a parametric instability will thus appear in the system as soon as \textit{one} of the harmonics of the modulation is close to twice the energy of \textit{any} of the (effective, time-averaged) Bogoliubov modes, $2\Eav(\qbf)$, i.e. $m\omega\!\approx\!2\Eav(\qbf)$ (resonance condition).  The later will result in a long-term explosion in the (stroboscopic) dynamics of the corresponding mode.
Assuming that the harmonics can be treated independently, the instability rate of a given mode $\qbf$ and a given harmonics can be analytically computed following the perturbative method developed in~\cite{landau1969,lellouch2017}, and the total instability rate and most unstable mode straightforwardly inferred from Eqs.~(\ref{eq:GammaDef}) and~(\ref{eq:qmumDef}).

\subsection{Explicit analytics in the strict resonant case $\omega=\Delta$}

As an illustration, we now consider the simple case $l=1$ (i.e. $\omega=\Delta$) and $\phi=0$ addressed in the main text. For the typical parameters used in Fig.~\ref{fig:NumRes}, one immediately sees that only the smaller harmonics of the modulation will allow the resonance condition to be fulfilled; this corresponds to the terms $m=\pm 1$ in Eq.~(\ref{eq:apphq}). We can thus disregard other harmonics, as well as the term $\hat W_\qbf(t)$ which has no long-term stroboscopic contribution~\footnote{Interestingly however, we find that this term can play a role for higher values of $l$, with non trivial effects arising from the non-commutation between the diagonal $\hat W_\qbf(t)$ and the off-diagonal matrix in Eqs.~\ref{eq:appBdGEPO}.}, and we get the effective model
	\begin{align}
		&i \partial_t \! \left( \begin{matrix} \tilde{u}'_{\qbf}  \\ \tilde{v}'_{\qbf} \end{matrix} \right)=\biggl[ \Eav(\qbf)\hat{\mathbf{1}}+\!\dfrac{\alpha_q\Eav(\qbf)}{2}\left( \begin{matrix} 0 & \cos(2\omega t)\mathrm e^{-2i\Eav(\qbf)t}        \\ -\cos(2\omega t)\mathrm e^{2i\Eav(\qbf)t} & 0 \end{matrix} \right)\biggr]\! \left( \begin{matrix} \tilde{u}'_{\qbf}  \\ \tilde{v}'_{\qbf} \end{matrix} \right),
		\label{eq:BdGEO2}
	\end{align}
	with
	\begin{equation}
		\alpha_\qbf=8J \mathcal{J}_{-1}(\alpha)\sin^2(q_x/2) \frac{g}{[\Eav(\qbf)]^2}
		\label{eq:appalpha}
	\end{equation}
	As shown in Refs.~\cite{lellouch2017,bukov2015}, Eq.~\ref{eq:BdGEO2} is indeed formally equivalent to the equations of motion describing a parametric oscillator of eigenfrequency $\Eav(\qbf)$, driven by a perturbation of amplitude $\alpha_\qbf$ and frequency $2\omega$. The quantity $\alpha_\qbf$ will thus be referred to as the perturbation's amplitude in the effective parametric oscillator model; it is the small parameter of the perturbative method used to compute the instability rates~\cite{landau1969,lellouch2017}.
At lowest order in $\alpha_q$, we find in this case
	\begin{equation}
		\Gamma =  \max_\qbf \Big[ g\dfrac{2J |\mathcal{J}_{-1}|(\alpha)\sin^2(q_x/2)}{\Eav(\qbf)} \sqrt{1-\left(\dfrac{[\omega-\Eav(\qbf)]\Eav(\qbf)}{2J \mathcal{J}_{-1}(\alpha)\sin^2(q_x/2) g}\right)^2} \Big].
		\label{eq:appGamma}
	\end{equation}
	In the purely 1D case (no transverse direction), this yields the analytical stability diagram on the bottom panel of Fig.~\ref{fig:NumRes} (although the latter is actually computed using the analytical expression at second order, which is an implicit equation). In that case, the most unstable mode is $q_x^\mathrm{mum}=\pi$ at the onset of instability.
	
	In the case where a continuous degree of freedom is present, we find two regimes : 
	\begin{itemize}
	\item[(i)] If $\omega\!>\!\sqrt{4\vert \Jeff \vert(4\vert\Jeff\vert+2g)}$, one finds
		\begin{equation}
		q_x^\mathrm{mum}=\pi; \quad (\qp^\mathrm{mum})^2/2m = \sqrt{g^2+\omega^2}-g-4|\Jeff|.
		\label{eq:appqmum1}
		\end{equation}
		\begin{equation}
		\Gamma= 2J\left|\mathcal{J}_1(K/\omega)\right|\dfrac{g}{\omega}.
		\label{eq:apptaux1}
		\end{equation}
		
		\item[(ii)] If $\omega\!<\!\sqrt{4\vert\Jeff\vert(4\vert\Jeff\vert+2g)}$, we find
		\begin{equation}
		q_x^\mathrm{mum}= 2 \arcsin \sqrt{\dfrac{\sqrt{g^2+\omega^2}-g}{4|\Jeff|}} ; \quad \qp^\mathrm{mum}=0.
		\label{eq:appqmum2}
		\end{equation}
		\begin{align}
		\Gamma=(\sqrt{g^2+\omega^2}-g) \dfrac{g}{\omega}.
		\label{eq:apptaux2}
		\end{align}
	\end{itemize}

\begin{thebibliography}{66}%
\makeatletter
\providecommand \@ifxundefined [1]{%
 \@ifx{#1\undefined}
}%
\providecommand \@ifnum [1]{%
 \ifnum #1\expandafter \@firstoftwo
 \else \expandafter \@secondoftwo
 \fi
}%
\providecommand \@ifx [1]{%
 \ifx #1\expandafter \@firstoftwo
 \else \expandafter \@secondoftwo
 \fi
}%
\providecommand \natexlab [1]{#1}%
\providecommand \enquote  [1]{``#1''}%
\providecommand \bibnamefont  [1]{#1}%
\providecommand \bibfnamefont [1]{#1}%
\providecommand \citenamefont [1]{#1}%
\providecommand \href@noop [0]{\@secondoftwo}%
\providecommand \href [0]{\begingroup \@sanitize@url \@href}%
\providecommand \@href[1]{\@@startlink{#1}\@@href}%
\providecommand \@@href[1]{\endgroup#1\@@endlink}%
\providecommand \@sanitize@url [0]{\catcode `\\12\catcode `\$12\catcode
  `\&12\catcode `\#12\catcode `\^12\catcode `\_12\catcode `\%12\relax}%
\providecommand \@@startlink[1]{}%
\providecommand \@@endlink[0]{}%
\providecommand \url  [0]{\begingroup\@sanitize@url \@url }%
\providecommand \@url [1]{\endgroup\@href {#1}{\urlprefix }}%
\providecommand \urlprefix  [0]{URL }%
\providecommand \Eprint [0]{\href }%
\providecommand \doibase [0]{http://dx.doi.org/}%
\providecommand \selectlanguage [0]{\@gobble}%
\providecommand \bibinfo  [0]{\@secondoftwo}%
\providecommand \bibfield  [0]{\@secondoftwo}%
\providecommand \translation [1]{[#1]}%
\providecommand \BibitemOpen [0]{}%
\providecommand \bibitemStop [0]{}%
\providecommand \bibitemNoStop [0]{.\EOS\space}%
\providecommand \EOS [0]{\spacefactor3000\relax}%
\providecommand \BibitemShut  [1]{\csname bibitem#1\endcsname}%
\let\auto@bib@innerbib\@empty
\bibitem [{\citenamefont {Oka}\ and\ \citenamefont {Aoki}(2009)}]{oka_09}%
  \BibitemOpen
  \bibfield  {author} {\bibinfo {author} {\bibfnamefont {T.}~\bibnamefont
  {Oka}}\ and\ \bibinfo {author} {\bibfnamefont {H.}~\bibnamefont {Aoki}},\
  }\bibfield  {title} {\enquote {\bibinfo {title} {Photovoltaic hall effect in
  graphene},}\ }\href@noop {} {\bibfield  {journal} {\bibinfo  {journal} {Phys.
  Rev. B}\ }\textbf {\bibinfo {volume} {79}},\ \bibinfo {pages} {081406}
  (\bibinfo {year} {2009})}\BibitemShut {NoStop}%
\bibitem [{\citenamefont {Kitagawa}\ \emph {et~al.}(2010)\citenamefont
  {Kitagawa}, \citenamefont {Berg}, \citenamefont {Rudner},\ and\ \citenamefont
  {Demler}}]{kitagawa2010}%
  \BibitemOpen
  \bibfield  {author} {\bibinfo {author} {\bibfnamefont {T.}~\bibnamefont
  {Kitagawa}}, \bibinfo {author} {\bibfnamefont {E.}~\bibnamefont {Berg}},
  \bibinfo {author} {\bibfnamefont {M.}~\bibnamefont {Rudner}}, \ and\ \bibinfo
  {author} {\bibfnamefont {E.}~\bibnamefont {Demler}},\ }\bibfield  {title}
  {\enquote {\bibinfo {title} {Topological characterization of
  periodically-driven quantum systems},}\ }\href@noop {} {\bibfield  {journal}
  {\bibinfo  {journal} {\Jprb}\ }\textbf {\bibinfo {volume} {82}},\ \bibinfo
  {pages} {235114} (\bibinfo {year} {2010})}\BibitemShut {NoStop}%
\bibitem [{\citenamefont {Lindner}\ \emph {et~al.}(2011)\citenamefont
  {Lindner}, \citenamefont {Refael},\ and\ \citenamefont
  {Galitski}}]{lindner2011}%
  \BibitemOpen
  \bibfield  {author} {\bibinfo {author} {\bibfnamefont {N.~H.}\ \bibnamefont
  {Lindner}}, \bibinfo {author} {\bibfnamefont {G.}~\bibnamefont {Refael}}, \
  and\ \bibinfo {author} {\bibfnamefont {V.}~\bibnamefont {Galitski}},\
  }\bibfield  {title} {\enquote {\bibinfo {title} {Floquet topological
  insulator in semiconductor quantum wells},}\ }\href@noop {} {\bibfield
  {journal} {\bibinfo  {journal} {\Jnatphys}\ }\textbf {\bibinfo {volume}
  {7}},\ \bibinfo {pages} {490--495} (\bibinfo {year} {2011})}\BibitemShut
  {NoStop}%
\bibitem [{\citenamefont {Cayssol}\ \emph {et~al.}(2013)\citenamefont
  {Cayssol}, \citenamefont {Dora}, \citenamefont {Simon},\ and\ \citenamefont
  {Moessner}}]{cayssol2013}%
  \BibitemOpen
  \bibfield  {author} {\bibinfo {author} {\bibfnamefont {J.}~\bibnamefont
  {Cayssol}}, \bibinfo {author} {\bibfnamefont {B.}~\bibnamefont {Dora}},
  \bibinfo {author} {\bibfnamefont {F.}~\bibnamefont {Simon}}, \ and\ \bibinfo
  {author} {\bibfnamefont {R.}~\bibnamefont {Moessner}},\ }\bibfield  {title}
  {\enquote {\bibinfo {title} {Floquet topological insulators},}\ }\href@noop
  {} {\bibfield  {journal} {\bibinfo  {journal} {Phys. Status Solidi RRL}\
  }\textbf {\bibinfo {volume} {7}},\ \bibinfo {pages} {101--108} (\bibinfo
  {year} {2013})}\BibitemShut {NoStop}%
\bibitem [{\citenamefont {Mitrano}\ \emph {et~al.}(2016)\citenamefont
  {Mitrano}, \citenamefont {Cantaluppi}, \citenamefont {Nicoletti},
  \citenamefont {Kaiser}, \citenamefont {Perucchi}, \citenamefont {Lupi},
  \citenamefont {Di~Pietro}, \citenamefont {Pontiroli}, \citenamefont
  {Ricc{\`o}}, \citenamefont {Clark} \emph {et~al.}}]{mitrano2016possible}%
  \BibitemOpen
  \bibfield  {author} {\bibinfo {author} {\bibfnamefont {Matteo}\ \bibnamefont
  {Mitrano}}, \bibinfo {author} {\bibfnamefont {Alice}\ \bibnamefont
  {Cantaluppi}}, \bibinfo {author} {\bibfnamefont {Daniele}\ \bibnamefont
  {Nicoletti}}, \bibinfo {author} {\bibfnamefont {Stefan}\ \bibnamefont
  {Kaiser}}, \bibinfo {author} {\bibfnamefont {A}~\bibnamefont {Perucchi}},
  \bibinfo {author} {\bibfnamefont {S}~\bibnamefont {Lupi}}, \bibinfo {author}
  {\bibfnamefont {P}~\bibnamefont {Di~Pietro}}, \bibinfo {author}
  {\bibfnamefont {D}~\bibnamefont {Pontiroli}}, \bibinfo {author}
  {\bibfnamefont {M}~\bibnamefont {Ricc{\`o}}}, \bibinfo {author}
  {\bibfnamefont {Stephen~R}\ \bibnamefont {Clark}},  \emph {et~al.},\
  }\bibfield  {title} {\enquote {\bibinfo {title} {Possible light-induced
  superconductivity in k3c60 at high temperature},}\ }\href@noop {} {\bibfield
  {journal} {\bibinfo  {journal} {Nature}\ }\textbf {\bibinfo {volume} {530}},\
  \bibinfo {pages} {461--464} (\bibinfo {year} {2016})}\BibitemShut {NoStop}%
\bibitem [{\citenamefont {Aidelsburger}\ \emph {et~al.}(2017)\citenamefont
  {Aidelsburger}, \citenamefont {Nascimbene},\ and\ \citenamefont
  {Goldman}}]{aidelsburger2017artificial}%
  \BibitemOpen
  \bibfield  {author} {\bibinfo {author} {\bibfnamefont {M}~\bibnamefont
  {Aidelsburger}}, \bibinfo {author} {\bibfnamefont {S}~\bibnamefont
  {Nascimbene}}, \ and\ \bibinfo {author} {\bibfnamefont {N}~\bibnamefont
  {Goldman}},\ }\bibfield  {title} {\enquote {\bibinfo {title} {Artificial
  gauge fields in materials and engineered systems},}\ }\href@noop {}
  {\bibfield  {journal} {\bibinfo  {journal} {arXiv preprint arXiv:1710.00851}\
  } (\bibinfo {year} {2017})}\BibitemShut {NoStop}%
\bibitem [{\citenamefont {Goldman}\ \emph {et~al.}(2014)\citenamefont
  {Goldman}, \citenamefont {Juzeliunas}, \citenamefont {Ohberg},\ and\
  \citenamefont {Spielman}}]{goldman2014a}%
  \BibitemOpen
  \bibfield  {author} {\bibinfo {author} {\bibfnamefont {N.}~\bibnamefont
  {Goldman}}, \bibinfo {author} {\bibfnamefont {G.}~\bibnamefont {Juzeliunas}},
  \bibinfo {author} {\bibfnamefont {P.}~\bibnamefont {Ohberg}}, \ and\ \bibinfo
  {author} {\bibfnamefont {I.~B.}\ \bibnamefont {Spielman}},\ }\bibfield
  {title} {\enquote {\bibinfo {title} {Light-induced gauge fields for ultracold
  atoms},}\ }\href@noop {} {\bibfield  {journal} {\bibinfo  {journal}
  {\JRepProgPhys}\ }\textbf {\bibinfo {volume} {77}} (\bibinfo {year}
  {2014})}\BibitemShut {NoStop}%
\bibitem [{\citenamefont {Goldman}\ and\ \citenamefont
  {Dalibard}(2014)}]{goldman2014b}%
  \BibitemOpen
  \bibfield  {author} {\bibinfo {author} {\bibfnamefont {N.}~\bibnamefont
  {Goldman}}\ and\ \bibinfo {author} {\bibfnamefont {J.}~\bibnamefont
  {Dalibard}},\ }\bibfield  {title} {\enquote {\bibinfo {title}
  {Periodically-driven quantum systems: Effective hamiltonians and engineered
  gauge fields},}\ }\href@noop {} {\bibfield  {journal} {\bibinfo  {journal}
  {\Jprx}\ }\textbf {\bibinfo {volume} {4}} (\bibinfo {year}
  {2014})}\BibitemShut {NoStop}%
\bibitem [{\citenamefont {Bukov}\ \emph
  {et~al.}(2015{\natexlab{a}})\citenamefont {Bukov}, \citenamefont
  {D'Alessio},\ and\ \citenamefont {Polkovnikov}}]{bukov2014}%
  \BibitemOpen
  \bibfield  {author} {\bibinfo {author} {\bibfnamefont {M.}~\bibnamefont
  {Bukov}}, \bibinfo {author} {\bibfnamefont {L.}~\bibnamefont {D'Alessio}}, \
  and\ \bibinfo {author} {\bibfnamefont {A.}~\bibnamefont {Polkovnikov}},\
  }\bibfield  {title} {\enquote {\bibinfo {title} {Universal high-frequency
  behaviour of periodically-driven quantum systems : from dynamical
  stabilization to floquet engineering},}\ }\href@noop {} {\bibfield  {journal}
  {\bibinfo  {journal} {\Jadvphys}\ }\textbf {\bibinfo {volume} {64}},\
  \bibinfo {pages} {139--226} (\bibinfo {year}
  {2015}{\natexlab{a}})}\BibitemShut {NoStop}%
\bibitem [{\citenamefont {Eckardt}(2016)}]{Eckardt:2016Review}%
  \BibitemOpen
  \bibfield  {author} {\bibinfo {author} {\bibfnamefont {A}~\bibnamefont
  {Eckardt}},\ }\href@noop {} {\enquote {\bibinfo {title} {{Atomic quantum
  gases in periodically driven optical lattices}},}\ } (\bibinfo {year}
  {2016}),\ \bibinfo {note} {arxiv:1606.08041v1}\BibitemShut {NoStop}%
\bibitem [{\citenamefont {Aidelsburger}\ \emph
  {et~al.}(2015{\natexlab{a}})\citenamefont {Aidelsburger}, \citenamefont
  {Lohse}, \citenamefont {Schweizer}, \citenamefont {Atala}, \citenamefont
  {Barreiro}, \citenamefont {Nascimbene}, \citenamefont {Cooper}, \citenamefont
  {Bloch},\ and\ \citenamefont {Goldman}}]{aidelsburger2014}%
  \BibitemOpen
  \bibfield  {author} {\bibinfo {author} {\bibfnamefont {M.}~\bibnamefont
  {Aidelsburger}}, \bibinfo {author} {\bibfnamefont {M.}~\bibnamefont {Lohse}},
  \bibinfo {author} {\bibfnamefont {C.}~\bibnamefont {Schweizer}}, \bibinfo
  {author} {\bibfnamefont {M.}~\bibnamefont {Atala}}, \bibinfo {author}
  {\bibfnamefont {J.}~\bibnamefont {Barreiro}}, \bibinfo {author}
  {\bibfnamefont {S.}~\bibnamefont {Nascimbene}}, \bibinfo {author}
  {\bibfnamefont {N.~R.}\ \bibnamefont {Cooper}}, \bibinfo {author}
  {\bibfnamefont {I.}~\bibnamefont {Bloch}}, \ and\ \bibinfo {author}
  {\bibfnamefont {N.}~\bibnamefont {Goldman}},\ }\bibfield  {title} {\enquote
  {\bibinfo {title} {Measuring the chern number of hofstadter bands with
  ultracold bosonic atoms},}\ }\href@noop {} {\bibfield  {journal} {\bibinfo
  {journal} {\Jnatphys}\ }\textbf {\bibinfo {volume} {11}} (\bibinfo {year}
  {2015}{\natexlab{a}})}\BibitemShut {NoStop}%
\bibitem [{\citenamefont {Jotzu}\ \emph {et~al.}(2014)\citenamefont {Jotzu},
  \citenamefont {Messer}, \citenamefont {Desbuquois}, \citenamefont {Lebrat},
  \citenamefont {Uehlinger}, \citenamefont {Greif},\ and\ \citenamefont
  {Esslinger}}]{jotzu2014}%
  \BibitemOpen
  \bibfield  {author} {\bibinfo {author} {\bibfnamefont {G.}~\bibnamefont
  {Jotzu}}, \bibinfo {author} {\bibfnamefont {M.}~\bibnamefont {Messer}},
  \bibinfo {author} {\bibfnamefont {R.}~\bibnamefont {Desbuquois}}, \bibinfo
  {author} {\bibfnamefont {M.}~\bibnamefont {Lebrat}}, \bibinfo {author}
  {\bibfnamefont {T.}~\bibnamefont {Uehlinger}}, \bibinfo {author}
  {\bibfnamefont {D.}~\bibnamefont {Greif}}, \ and\ \bibinfo {author}
  {\bibfnamefont {T.}~\bibnamefont {Esslinger}},\ }\bibfield  {title} {\enquote
  {\bibinfo {title} {Experimental realisation of the topological haldane
  model},}\ }\href@noop {} {\bibfield  {journal} {\bibinfo  {journal}
  {\Jnature}\ }\textbf {\bibinfo {volume} {515}},\ \bibinfo {pages} {237--240}
  (\bibinfo {year} {2014})}\BibitemShut {NoStop}%
\bibitem [{\citenamefont {Goldman}\ \emph {et~al.}(2016)\citenamefont
  {Goldman}, \citenamefont {Budich},\ and\ \citenamefont
  {Zoller}}]{Goldman:2016Review}%
  \BibitemOpen
  \bibfield  {author} {\bibinfo {author} {\bibfnamefont {N.}~\bibnamefont
  {Goldman}}, \bibinfo {author} {\bibfnamefont {J.C.}\ \bibnamefont {Budich}},
  \ and\ \bibinfo {author} {\bibfnamefont {P.}~\bibnamefont {Zoller}},\
  }\bibfield  {title} {\enquote {\bibinfo {title} {{Topological quantum matter
  with ultracold gases in optical lattices}},}\ }\href@noop {} {\bibfield
  {journal} {\bibinfo  {journal} {Nature Physics}\ }\textbf {\bibinfo {volume}
  {12}},\ \bibinfo {pages} {639--645} (\bibinfo {year} {2016})}\BibitemShut
  {NoStop}%
\bibitem [{\citenamefont {Eckardt}\ \emph {et~al.}(2010)\citenamefont
  {Eckardt}, \citenamefont {Hauke}, \citenamefont {Soltan-Panahi},
  \citenamefont {Becker}, \citenamefont {Sengtock},\ and\ \citenamefont
  {Lewenstein}}]{eckardt2010}%
  \BibitemOpen
  \bibfield  {author} {\bibinfo {author} {\bibfnamefont {A.}~\bibnamefont
  {Eckardt}}, \bibinfo {author} {\bibfnamefont {P.}~\bibnamefont {Hauke}},
  \bibinfo {author} {\bibfnamefont {P.}~\bibnamefont {Soltan-Panahi}}, \bibinfo
  {author} {\bibfnamefont {C.}~\bibnamefont {Becker}}, \bibinfo {author}
  {\bibfnamefont {K.}~\bibnamefont {Sengtock}}, \ and\ \bibinfo {author}
  {\bibfnamefont {M.}~\bibnamefont {Lewenstein}},\ }\bibfield  {title}
  {\enquote {\bibinfo {title} {Frustrated quantum antiferromagnetism with
  ultracold bosons in a triangular lattice},}\ }\href@noop {} {\bibfield
  {journal} {\bibinfo  {journal} {\Jepl}\ }\textbf {\bibinfo {volume} {89}},\
  \bibinfo {pages} {10010} (\bibinfo {year} {2010})}\BibitemShut {NoStop}%
\bibitem [{\citenamefont {Struck}\ \emph {et~al.}(2011)\citenamefont {Struck},
  \citenamefont {Olschlager}, \citenamefont {{Le Targat}}, \citenamefont
  {Soltan-Panahi}, \citenamefont {Eckardt}, \citenamefont {Lewenstein},
  \citenamefont {Windpassinger},\ and\ \citenamefont {Sengstock}}]{struck2011}%
  \BibitemOpen
  \bibfield  {author} {\bibinfo {author} {\bibfnamefont {J.}~\bibnamefont
  {Struck}}, \bibinfo {author} {\bibfnamefont {C.}~\bibnamefont {Olschlager}},
  \bibinfo {author} {\bibfnamefont {R.}~\bibnamefont {{Le Targat}}}, \bibinfo
  {author} {\bibfnamefont {P.}~\bibnamefont {Soltan-Panahi}}, \bibinfo {author}
  {\bibfnamefont {A.}~\bibnamefont {Eckardt}}, \bibinfo {author} {\bibfnamefont
  {M.}~\bibnamefont {Lewenstein}}, \bibinfo {author} {\bibfnamefont
  {P.}~\bibnamefont {Windpassinger}}, \ and\ \bibinfo {author} {\bibfnamefont
  {K.}~\bibnamefont {Sengstock}},\ }\bibfield  {title} {\enquote {\bibinfo
  {title} {Quantum simulation of frustrated classical magnetism in triangular
  optical lattices},}\ }\href@noop {} {\bibfield  {journal} {\bibinfo
  {journal} {\Jscience}\ }\textbf {\bibinfo {volume} {333}},\ \bibinfo {pages}
  {996--999} (\bibinfo {year} {2011})}\BibitemShut {NoStop}%
\bibitem [{\citenamefont {{G{\"o}rg}}\ \emph {et~al.}(2017)\citenamefont
  {{G{\"o}rg}}, \citenamefont {{Messer}}, \citenamefont {{Sandholzer}},
  \citenamefont {{Jotzu}}, \citenamefont {{Desbuquois}},\ and\ \citenamefont
  {{Esslinger}}}]{Gorg_2017}%
  \BibitemOpen
  \bibfield  {author} {\bibinfo {author} {\bibfnamefont {F.}~\bibnamefont
  {{G{\"o}rg}}}, \bibinfo {author} {\bibfnamefont {M.}~\bibnamefont
  {{Messer}}}, \bibinfo {author} {\bibfnamefont {K.}~\bibnamefont
  {{Sandholzer}}}, \bibinfo {author} {\bibfnamefont {G.}~\bibnamefont
  {{Jotzu}}}, \bibinfo {author} {\bibfnamefont {R.}~\bibnamefont
  {{Desbuquois}}}, \ and\ \bibinfo {author} {\bibfnamefont {T.}~\bibnamefont
  {{Esslinger}}},\ }\bibfield  {title} {\enquote {\bibinfo {title}
  {{Enhancement and sign reversal of magnetic correlations in a driven quantum
  many-body system}},}\ }\href@noop {} {\bibfield  {journal} {\bibinfo
  {journal} {ArXiv e-prints}\ } (\bibinfo {year} {2017})},\ \Eprint
  {http://arxiv.org/abs/1708.06751} {arXiv:1708.06751 [cond-mat.quant-gas]}
  \BibitemShut {NoStop}%
\bibitem [{\citenamefont {Aidelsburger}\ \emph {et~al.}(2011)\citenamefont
  {Aidelsburger}, \citenamefont {Atala}, \citenamefont {Nascimbene},
  \citenamefont {Trotzky}, \citenamefont {Chen},\ and\ \citenamefont
  {Bloch}}]{aidelsburger2011}%
  \BibitemOpen
  \bibfield  {author} {\bibinfo {author} {\bibfnamefont {M.}~\bibnamefont
  {Aidelsburger}}, \bibinfo {author} {\bibfnamefont {M.}~\bibnamefont {Atala}},
  \bibinfo {author} {\bibfnamefont {S.}~\bibnamefont {Nascimbene}}, \bibinfo
  {author} {\bibfnamefont {S.}~\bibnamefont {Trotzky}}, \bibinfo {author}
  {\bibfnamefont {Y.}~\bibnamefont {Chen}}, \ and\ \bibinfo {author}
  {\bibfnamefont {I.}~\bibnamefont {Bloch}},\ }\bibfield  {title} {\enquote
  {\bibinfo {title} {Experimental realization of strong effective magnetic
  fields in an optical lattice},}\ }\href@noop {} {\bibfield  {journal}
  {\bibinfo  {journal} {\Jprl}\ }\textbf {\bibinfo {volume} {107}} (\bibinfo
  {year} {2011})}\BibitemShut {NoStop}%
\bibitem [{\citenamefont {Aidelsburger}\ \emph {et~al.}(2013)\citenamefont
  {Aidelsburger}, \citenamefont {Atala}, \citenamefont {Lohse}, \citenamefont
  {Barreiro}, \citenamefont {Paredes},\ and\ \citenamefont
  {Bloch}}]{aidelsburger2013}%
  \BibitemOpen
  \bibfield  {author} {\bibinfo {author} {\bibfnamefont {M.}~\bibnamefont
  {Aidelsburger}}, \bibinfo {author} {\bibfnamefont {M.}~\bibnamefont {Atala}},
  \bibinfo {author} {\bibfnamefont {M.}~\bibnamefont {Lohse}}, \bibinfo
  {author} {\bibfnamefont {J.}~\bibnamefont {Barreiro}}, \bibinfo {author}
  {\bibfnamefont {B.}~\bibnamefont {Paredes}}, \ and\ \bibinfo {author}
  {\bibfnamefont {I.}~\bibnamefont {Bloch}},\ }\bibfield  {title} {\enquote
  {\bibinfo {title} {Realization of the hofstadter hamiltonian with ultracold
  atoms in optical lattices},}\ }\href@noop {} {\bibfield  {journal} {\bibinfo
  {journal} {\Jprl}\ }\textbf {\bibinfo {volume} {111}},\ \bibinfo {pages}
  {185301} (\bibinfo {year} {2013})}\BibitemShut {NoStop}%
\bibitem [{\citenamefont {Miyake}\ \emph {et~al.}(2013)\citenamefont {Miyake},
  \citenamefont {Siviloglou}, \citenamefont {Kennedy}, \citenamefont {Burton},\
  and\ \citenamefont {Ketterle}}]{miyake2013}%
  \BibitemOpen
  \bibfield  {author} {\bibinfo {author} {\bibfnamefont {H.}~\bibnamefont
  {Miyake}}, \bibinfo {author} {\bibfnamefont {G.~A.}\ \bibnamefont
  {Siviloglou}}, \bibinfo {author} {\bibfnamefont {C.~J.}\ \bibnamefont
  {Kennedy}}, \bibinfo {author} {\bibfnamefont {W.~C.}\ \bibnamefont {Burton}},
  \ and\ \bibinfo {author} {\bibfnamefont {W.}~\bibnamefont {Ketterle}},\
  }\bibfield  {title} {\enquote {\bibinfo {title} {Realizing the harper
  hamiltonian with laser-assisted tunneling in optical lattices},}\ }\href@noop
  {} {\bibfield  {journal} {\bibinfo  {journal} {\Jprl}\ }\textbf {\bibinfo
  {volume} {111}},\ \bibinfo {pages} {185302} (\bibinfo {year}
  {2013})}\BibitemShut {NoStop}%
\bibitem [{\citenamefont {Kennedy}\ \emph {et~al.}(2015)\citenamefont
  {Kennedy}, \citenamefont {Burton}, \citenamefont {Chung},\ and\ \citenamefont
  {Ketterle}}]{kennedy2015}%
  \BibitemOpen
  \bibfield  {author} {\bibinfo {author} {\bibfnamefont {C.~J.}\ \bibnamefont
  {Kennedy}}, \bibinfo {author} {\bibfnamefont {W.~C.}\ \bibnamefont {Burton}},
  \bibinfo {author} {\bibfnamefont {W.~C.}\ \bibnamefont {Chung}}, \ and\
  \bibinfo {author} {\bibfnamefont {W.}~\bibnamefont {Ketterle}},\ }\bibfield
  {title} {\enquote {\bibinfo {title} {Observation of bose-einstein
  condensation in a strong synthetic magnetic field, arxiv:150308243},}\
  }\href@noop {} {\bibfield  {journal} {\bibinfo  {journal} {Nature Physics}\
  }\textbf {\bibinfo {volume} {11}},\ \bibinfo {pages} {859--864} (\bibinfo
  {year} {2015})}\BibitemShut {NoStop}%
\bibitem [{\citenamefont {Tai}\ \emph {et~al.}(2017)\citenamefont {Tai},
  \citenamefont {Lukin}, \citenamefont {Rispoli}, \citenamefont {Schittko},
  \citenamefont {Menke}, \citenamefont {Borgnia}, \citenamefont {Preiss},
  \citenamefont {Grusdt}, \citenamefont {Kaufman},\ and\ \citenamefont
  {Greiner}}]{tai2017microscopy}%
  \BibitemOpen
  \bibfield  {author} {\bibinfo {author} {\bibfnamefont {M~Eric}\ \bibnamefont
  {Tai}}, \bibinfo {author} {\bibfnamefont {Alexander}\ \bibnamefont {Lukin}},
  \bibinfo {author} {\bibfnamefont {Matthew}\ \bibnamefont {Rispoli}}, \bibinfo
  {author} {\bibfnamefont {Robert}\ \bibnamefont {Schittko}}, \bibinfo {author}
  {\bibfnamefont {Tim}\ \bibnamefont {Menke}}, \bibinfo {author} {\bibfnamefont
  {Dan}\ \bibnamefont {Borgnia}}, \bibinfo {author} {\bibfnamefont {Philipp~M}\
  \bibnamefont {Preiss}}, \bibinfo {author} {\bibfnamefont {Fabian}\
  \bibnamefont {Grusdt}}, \bibinfo {author} {\bibfnamefont {Adam~M}\
  \bibnamefont {Kaufman}}, \ and\ \bibinfo {author} {\bibfnamefont {Markus}\
  \bibnamefont {Greiner}},\ }\bibfield  {title} {\enquote {\bibinfo {title}
  {Microscopy of the interacting harper--hofstadter model in the two-body
  limit},}\ }\href@noop {} {\bibfield  {journal} {\bibinfo  {journal} {Nature}\
  }\textbf {\bibinfo {volume} {546}},\ \bibinfo {pages} {519--523} (\bibinfo
  {year} {2017})}\BibitemShut {NoStop}%
\bibitem [{\citenamefont {Tarnowski}\ \emph {et~al.}(2017)\citenamefont
  {Tarnowski}, \citenamefont {{\"U}nal}, \citenamefont {Fl{\"a}schner},
  \citenamefont {Rem}, \citenamefont {Eckardt}, \citenamefont {Sengstock},\
  and\ \citenamefont {Weitenberg}}]{tarnowski2017characterizing}%
  \BibitemOpen
  \bibfield  {author} {\bibinfo {author} {\bibfnamefont {Matthias}\
  \bibnamefont {Tarnowski}}, \bibinfo {author} {\bibfnamefont {F~Nur}\
  \bibnamefont {{\"U}nal}}, \bibinfo {author} {\bibfnamefont {Nick}\
  \bibnamefont {Fl{\"a}schner}}, \bibinfo {author} {\bibfnamefont {Benno~S}\
  \bibnamefont {Rem}}, \bibinfo {author} {\bibfnamefont {Andr{\'e}}\
  \bibnamefont {Eckardt}}, \bibinfo {author} {\bibfnamefont {Klaus}\
  \bibnamefont {Sengstock}}, \ and\ \bibinfo {author} {\bibfnamefont
  {Christof}\ \bibnamefont {Weitenberg}},\ }\bibfield  {title} {\enquote
  {\bibinfo {title} {Characterizing topology by dynamics: Chern number from
  linking number},}\ }\href@noop {} {\bibfield  {journal} {\bibinfo  {journal}
  {arXiv preprint arXiv:1709.01046}\ } (\bibinfo {year} {2017})}\BibitemShut
  {NoStop}%
\bibitem [{\citenamefont {Kolovsky}(2011)}]{kolovsky_11}%
  \BibitemOpen
  \bibfield  {author} {\bibinfo {author} {\bibfnamefont {A.}~\bibnamefont
  {Kolovsky}},\ }\href@noop {} {\bibfield  {journal} {\bibinfo  {journal}
  {EPL}\ }\textbf {\bibinfo {volume} {93}},\ \bibinfo {pages} {20003} (\bibinfo
  {year} {2011})}\BibitemShut {NoStop}%
\bibitem [{\citenamefont {Bermudez}\ \emph {et~al.}(2011)\citenamefont
  {Bermudez}, \citenamefont {Schaetz},\ and\ \citenamefont
  {Porras}}]{bermudez2011synthetic}%
  \BibitemOpen
  \bibfield  {author} {\bibinfo {author} {\bibfnamefont {Alejandro}\
  \bibnamefont {Bermudez}}, \bibinfo {author} {\bibfnamefont {Tobias}\
  \bibnamefont {Schaetz}}, \ and\ \bibinfo {author} {\bibfnamefont {Diego}\
  \bibnamefont {Porras}},\ }\bibfield  {title} {\enquote {\bibinfo {title}
  {Synthetic gauge fields for vibrational excitations of trapped ions},}\
  }\href@noop {} {\bibfield  {journal} {\bibinfo  {journal} {Physical review
  letters}\ }\textbf {\bibinfo {volume} {107}},\ \bibinfo {pages} {150501}
  (\bibinfo {year} {2011})}\BibitemShut {NoStop}%
\bibitem [{\citenamefont {Hauke}\ \emph {et~al.}(2012)\citenamefont {Hauke},
  \citenamefont {Tieleman}, \citenamefont {Celi}, \citenamefont
  {{\"O}lschl{\"a}ger}, \citenamefont {Simonet}, \citenamefont {Struck},
  \citenamefont {Weinberg}, \citenamefont {Windpassinger}, \citenamefont
  {Sengstock}, \citenamefont {Lewenstein},\ and\ \citenamefont
  {Eckardt}}]{hauke_12}%
  \BibitemOpen
  \bibfield  {author} {\bibinfo {author} {\bibfnamefont {Philipp}\ \bibnamefont
  {Hauke}}, \bibinfo {author} {\bibfnamefont {Olivier}\ \bibnamefont
  {Tieleman}}, \bibinfo {author} {\bibfnamefont {Alessio}\ \bibnamefont
  {Celi}}, \bibinfo {author} {\bibfnamefont {Christoph}\ \bibnamefont
  {{\"O}lschl{\"a}ger}}, \bibinfo {author} {\bibfnamefont {Juliette}\
  \bibnamefont {Simonet}}, \bibinfo {author} {\bibfnamefont {Julian}\
  \bibnamefont {Struck}}, \bibinfo {author} {\bibfnamefont {Malte}\
  \bibnamefont {Weinberg}}, \bibinfo {author} {\bibfnamefont {Patrick}\
  \bibnamefont {Windpassinger}}, \bibinfo {author} {\bibfnamefont {Klaus}\
  \bibnamefont {Sengstock}}, \bibinfo {author} {\bibfnamefont {Maciej}\
  \bibnamefont {Lewenstein}}, \ and\ \bibinfo {author} {\bibfnamefont
  {Andr{\'e}}\ \bibnamefont {Eckardt}},\ }\href@noop {} {\bibfield  {journal}
  {\bibinfo  {journal} {Phys. Rev. Lett.}\ }\textbf {\bibinfo {volume} {109}},\
  \bibinfo {pages} {145301} (\bibinfo {year} {2012})}\BibitemShut {NoStop}%
\bibitem [{\citenamefont {Goldman}\ \emph {et~al.}(2015)\citenamefont
  {Goldman}, \citenamefont {Dalibard}, \citenamefont {Aidelsburger},\ and\
  \citenamefont {Cooper}}]{goldman2015a}%
  \BibitemOpen
  \bibfield  {author} {\bibinfo {author} {\bibfnamefont {N.}~\bibnamefont
  {Goldman}}, \bibinfo {author} {\bibfnamefont {J.}~\bibnamefont {Dalibard}},
  \bibinfo {author} {\bibfnamefont {M.}~\bibnamefont {Aidelsburger}}, \ and\
  \bibinfo {author} {\bibfnamefont {N.}~\bibnamefont {Cooper}},\ }\bibfield
  {title} {\enquote {\bibinfo {title} {Periodically-driven quantum matter: the
  case of resonant modulations},}\ }\href@noop {} {\bibfield  {journal}
  {\bibinfo  {journal} {\Jpra}\ }\textbf {\bibinfo {volume} {91}} (\bibinfo
  {year} {2015})}\BibitemShut {NoStop}%
\bibitem [{\citenamefont {Creffield}\ \emph {et~al.}(2016)\citenamefont
  {Creffield}, \citenamefont {Pieplow}, \citenamefont {Sols},\ and\
  \citenamefont {Goldman}}]{creffield2016}%
  \BibitemOpen
  \bibfield  {author} {\bibinfo {author} {\bibfnamefont {C.~E.}\ \bibnamefont
  {Creffield}}, \bibinfo {author} {\bibfnamefont {G.}~\bibnamefont {Pieplow}},
  \bibinfo {author} {\bibfnamefont {F.}~\bibnamefont {Sols}}, \ and\ \bibinfo
  {author} {\bibfnamefont {N.}~\bibnamefont {Goldman}},\ }\bibfield  {title}
  {\enquote {\bibinfo {title} {Realization of uniform synthetic magnetic fields
  by periodically shaking an optical square lattice},}\ }\href@noop {}
  {\bibfield  {journal} {\bibinfo  {journal} {\Jnjp}\ }\textbf {\bibinfo
  {volume} {18}},\ \bibinfo {pages} {093013} (\bibinfo {year}
  {2016})}\BibitemShut {NoStop}%
\bibitem [{\citenamefont {Price}\ \emph {et~al.}(2017)\citenamefont {Price},
  \citenamefont {Ozawa},\ and\ \citenamefont {Goldman}}]{price2017synthetic}%
  \BibitemOpen
  \bibfield  {author} {\bibinfo {author} {\bibfnamefont {Hannah~M}\
  \bibnamefont {Price}}, \bibinfo {author} {\bibfnamefont {Tomoki}\
  \bibnamefont {Ozawa}}, \ and\ \bibinfo {author} {\bibfnamefont {Nathan}\
  \bibnamefont {Goldman}},\ }\bibfield  {title} {\enquote {\bibinfo {title}
  {Synthetic dimensions for cold atoms from shaking a harmonic trap},}\
  }\href@noop {} {\bibfield  {journal} {\bibinfo  {journal} {Physical Review
  A}\ }\textbf {\bibinfo {volume} {95}},\ \bibinfo {pages} {023607} (\bibinfo
  {year} {2017})}\BibitemShut {NoStop}%
\bibitem [{\citenamefont {Sias}\ \emph {et~al.}(2008)\citenamefont {Sias},
  \citenamefont {Lignier}, \citenamefont {Singh}, \citenamefont {Zenesini},
  \citenamefont {Ciampini}, \citenamefont {Morsch},\ and\ \citenamefont
  {Arimondo}}]{sias2008}%
  \BibitemOpen
  \bibfield  {author} {\bibinfo {author} {\bibfnamefont {C.}~\bibnamefont
  {Sias}}, \bibinfo {author} {\bibfnamefont {H.}~\bibnamefont {Lignier}},
  \bibinfo {author} {\bibfnamefont {Y.}~\bibnamefont {Singh}}, \bibinfo
  {author} {\bibfnamefont {A.}~\bibnamefont {Zenesini}}, \bibinfo {author}
  {\bibfnamefont {D.}~\bibnamefont {Ciampini}}, \bibinfo {author}
  {\bibfnamefont {O.}~\bibnamefont {Morsch}}, \ and\ \bibinfo {author}
  {\bibfnamefont {E.}~\bibnamefont {Arimondo}},\ }\bibfield  {title} {\enquote
  {\bibinfo {title} {Observation of photon-assisted tunneling in optical
  lattices},}\ }\href@noop {} {\bibfield  {journal} {\bibinfo  {journal}
  {\Jprl}\ }\textbf {\bibinfo {volume} {100}},\ \bibinfo {pages} {040404}
  (\bibinfo {year} {2008})}\BibitemShut {NoStop}%
\bibitem [{\citenamefont {Mukherjee}\ \emph {et~al.}(2015)\citenamefont
  {Mukherjee}, \citenamefont {Spracklen}, \citenamefont {Choudhury},
  \citenamefont {Goldman}, \citenamefont {{\"O}hberg}, \citenamefont
  {Andersson},\ and\ \citenamefont {Thomson}}]{mukherjee2015modulation}%
  \BibitemOpen
  \bibfield  {author} {\bibinfo {author} {\bibfnamefont {Sebabrata}\
  \bibnamefont {Mukherjee}}, \bibinfo {author} {\bibfnamefont {Alexander}\
  \bibnamefont {Spracklen}}, \bibinfo {author} {\bibfnamefont {Debaditya}\
  \bibnamefont {Choudhury}}, \bibinfo {author} {\bibfnamefont {Nathan}\
  \bibnamefont {Goldman}}, \bibinfo {author} {\bibfnamefont {Patrik}\
  \bibnamefont {{\"O}hberg}}, \bibinfo {author} {\bibfnamefont {Erika}\
  \bibnamefont {Andersson}}, \ and\ \bibinfo {author} {\bibfnamefont
  {Robert~R}\ \bibnamefont {Thomson}},\ }\bibfield  {title} {\enquote {\bibinfo
  {title} {Modulation-assisted tunneling in laser-fabricated photonic
  wannier--stark ladders},}\ }\href@noop {} {\bibfield  {journal} {\bibinfo
  {journal} {New Journal of Physics}\ }\textbf {\bibinfo {volume} {17}},\
  \bibinfo {pages} {115002} (\bibinfo {year} {2015})}\BibitemShut {NoStop}%
\bibitem [{\citenamefont {Aidelsburger}\ \emph
  {et~al.}(2015{\natexlab{b}})\citenamefont {Aidelsburger}, \citenamefont
  {Lohse}, \citenamefont {Schweizer}, \citenamefont {Atala}, \citenamefont
  {Barreiro}, \citenamefont {Nascimb{\`e}ne}, \citenamefont {Cooper},
  \citenamefont {Bloch},\ and\ \citenamefont {Goldman}}]{aidelsburger_14}%
  \BibitemOpen
  \bibfield  {author} {\bibinfo {author} {\bibfnamefont {M.}~\bibnamefont
  {Aidelsburger}}, \bibinfo {author} {\bibfnamefont {M.}~\bibnamefont {Lohse}},
  \bibinfo {author} {\bibfnamefont {C.}~\bibnamefont {Schweizer}}, \bibinfo
  {author} {\bibfnamefont {M}~\bibnamefont {Atala}}, \bibinfo {author}
  {\bibfnamefont {J.~T.}\ \bibnamefont {Barreiro}}, \bibinfo {author}
  {\bibfnamefont {S.}~\bibnamefont {Nascimb{\`e}ne}}, \bibinfo {author}
  {\bibfnamefont {N.~R.}\ \bibnamefont {Cooper}}, \bibinfo {author}
  {\bibfnamefont {I.}~\bibnamefont {Bloch}}, \ and\ \bibinfo {author}
  {\bibfnamefont {N.}~\bibnamefont {Goldman}},\ }\href@noop {} {\bibfield
  {journal} {\bibinfo  {journal} {Nature Physics}\ }\textbf {\bibinfo {volume}
  {11}},\ \bibinfo {pages} {162--166} (\bibinfo {year}
  {2015}{\natexlab{b}})}\BibitemShut {NoStop}%
\bibitem [{\citenamefont {Hofstadter}(1976)}]{hofstadter1976}%
  \BibitemOpen
  \bibfield  {author} {\bibinfo {author} {\bibfnamefont {Douglas~R.}\
  \bibnamefont {Hofstadter}},\ }\bibfield  {title} {\enquote {\bibinfo {title}
  {Energy levels and wave functions of {B}loch electrons in rational and
  irrational magnetic fields},}\ }\href@noop {} {\bibfield  {journal} {\bibinfo
   {journal} {\Jprb}\ }\textbf {\bibinfo {volume} {14}},\ \bibinfo {pages}
  {2239--2249} (\bibinfo {year} {1976})}\BibitemShut {NoStop}%
\bibitem [{\citenamefont {Kohmoto}(1989)}]{kohmoto1989zero}%
  \BibitemOpen
  \bibfield  {author} {\bibinfo {author} {\bibfnamefont {Mahito}\ \bibnamefont
  {Kohmoto}},\ }\bibfield  {title} {\enquote {\bibinfo {title} {Zero modes and
  the quantized hall conductance of the two-dimensional lattice in a magnetic
  field},}\ }\href@noop {} {\bibfield  {journal} {\bibinfo  {journal} {Physical
  Review B}\ }\textbf {\bibinfo {volume} {39}},\ \bibinfo {pages} {11943}
  (\bibinfo {year} {1989})}\BibitemShut {NoStop}%
\bibitem [{\citenamefont {M{\"o}ller}\ and\ \citenamefont
  {Cooper}(2015)}]{moller2015fractional}%
  \BibitemOpen
  \bibfield  {author} {\bibinfo {author} {\bibfnamefont {Gunnar}\ \bibnamefont
  {M{\"o}ller}}\ and\ \bibinfo {author} {\bibfnamefont {Nigel~R}\ \bibnamefont
  {Cooper}},\ }\bibfield  {title} {\enquote {\bibinfo {title} {Fractional chern
  insulators in harper-hofstadter bands with higher chern number},}\
  }\href@noop {} {\bibfield  {journal} {\bibinfo  {journal} {Physical review
  letters}\ }\textbf {\bibinfo {volume} {115}},\ \bibinfo {pages} {126401}
  (\bibinfo {year} {2015})}\BibitemShut {NoStop}%
\bibitem [{\citenamefont {Regnault}\ and\ \citenamefont
  {Bernevig}(2011)}]{regnault2011fractional}%
  \BibitemOpen
  \bibfield  {author} {\bibinfo {author} {\bibfnamefont {Nicolas}\ \bibnamefont
  {Regnault}}\ and\ \bibinfo {author} {\bibfnamefont {B~Andrei}\ \bibnamefont
  {Bernevig}},\ }\bibfield  {title} {\enquote {\bibinfo {title} {Fractional
  chern insulator},}\ }\href@noop {} {\bibfield  {journal} {\bibinfo  {journal}
  {Physical Review X}\ }\textbf {\bibinfo {volume} {1}},\ \bibinfo {pages}
  {021014} (\bibinfo {year} {2011})}\BibitemShut {NoStop}%
\bibitem [{\citenamefont {Sterdyniak}\ \emph {et~al.}(2013)\citenamefont
  {Sterdyniak}, \citenamefont {Repellin}, \citenamefont {Bernevig},\ and\
  \citenamefont {Regnault}}]{sterdyniak2013series}%
  \BibitemOpen
  \bibfield  {author} {\bibinfo {author} {\bibfnamefont {Antoine}\ \bibnamefont
  {Sterdyniak}}, \bibinfo {author} {\bibfnamefont {Cecile}\ \bibnamefont
  {Repellin}}, \bibinfo {author} {\bibfnamefont {B~Andrei}\ \bibnamefont
  {Bernevig}}, \ and\ \bibinfo {author} {\bibfnamefont {Nicolas}\ \bibnamefont
  {Regnault}},\ }\bibfield  {title} {\enquote {\bibinfo {title} {Series of
  abelian and non-abelian states in c> 1 fractional chern insulators},}\
  }\href@noop {} {\bibfield  {journal} {\bibinfo  {journal} {Physical Review
  B}\ }\textbf {\bibinfo {volume} {87}},\ \bibinfo {pages} {205137} (\bibinfo
  {year} {2013})}\BibitemShut {NoStop}%
\bibitem [{\citenamefont {Bergholtz}\ and\ \citenamefont
  {Liu}(2013)}]{bergholtz2013topological}%
  \BibitemOpen
  \bibfield  {author} {\bibinfo {author} {\bibfnamefont {Emil~J}\ \bibnamefont
  {Bergholtz}}\ and\ \bibinfo {author} {\bibfnamefont {Zhao}\ \bibnamefont
  {Liu}},\ }\bibfield  {title} {\enquote {\bibinfo {title} {Topological flat
  band models and fractional chern insulators},}\ }\href@noop {} {\bibfield
  {journal} {\bibinfo  {journal} {International Journal of Modern Physics B}\
  }\textbf {\bibinfo {volume} {27}},\ \bibinfo {pages} {1330017} (\bibinfo
  {year} {2013})}\BibitemShut {NoStop}%
\bibitem [{\citenamefont {Grushin}\ \emph {et~al.}(2014)\citenamefont
  {Grushin}, \citenamefont {G{\'o}mez-Le{\'o}n},\ and\ \citenamefont
  {Neupert}}]{grushin_14}%
  \BibitemOpen
  \bibfield  {author} {\bibinfo {author} {\bibfnamefont {A.~G.}\ \bibnamefont
  {Grushin}}, \bibinfo {author} {\bibfnamefont {A.}~\bibnamefont
  {G{\'o}mez-Le{\'o}n}}, \ and\ \bibinfo {author} {\bibfnamefont
  {T.}~\bibnamefont {Neupert}},\ }\bibfield  {title} {\enquote {\bibinfo
  {title} {Floquet fractional chern insulators},}\ }\href@noop {} {\bibfield
  {journal} {\bibinfo  {journal} {Phys. Rev. Lett.}\ }\textbf {\bibinfo
  {volume} {112}},\ \bibinfo {pages} {156801} (\bibinfo {year}
  {2014})}\BibitemShut {NoStop}%
\bibitem [{\citenamefont {Reitter}\ \emph {et~al.}(2017)\citenamefont
  {Reitter}, \citenamefont {Näger}, \citenamefont {Wintersperger},
  \citenamefont {Sträter}, \citenamefont {Bloch}, \citenamefont {Eckardt},\
  and\ \citenamefont {Schneider}}]{reitter2017}%
  \BibitemOpen
  \bibfield  {author} {\bibinfo {author} {\bibfnamefont {M.}~\bibnamefont
  {Reitter}}, \bibinfo {author} {\bibfnamefont {J.}~\bibnamefont {Näger}},
  \bibinfo {author} {\bibfnamefont {K.}~\bibnamefont {Wintersperger}}, \bibinfo
  {author} {\bibfnamefont {C.}~\bibnamefont {Sträter}}, \bibinfo {author}
  {\bibfnamefont {I.}~\bibnamefont {Bloch}}, \bibinfo {author} {\bibfnamefont
  {A.}~\bibnamefont {Eckardt}}, \ and\ \bibinfo {author} {\bibfnamefont
  {U.}~\bibnamefont {Schneider}},\ }\href@noop {} {\enquote {\bibinfo {title}
  {Interaction dependent heating and atom loss in a periodically driven optical
  lattice},}\ } (\bibinfo {year} {2017}),\ \bibinfo {note}
  {arxiv:1706.04819}\BibitemShut {NoStop}%
\bibitem [{\citenamefont {Lellouch}\ \emph {et~al.}(2017)\citenamefont
  {Lellouch}, \citenamefont {Bukov}, \citenamefont {Demler},\ and\
  \citenamefont {Goldman}}]{lellouch2017}%
  \BibitemOpen
  \bibfield  {author} {\bibinfo {author} {\bibfnamefont {S.}~\bibnamefont
  {Lellouch}}, \bibinfo {author} {\bibfnamefont {M.}~\bibnamefont {Bukov}},
  \bibinfo {author} {\bibfnamefont {E.}~\bibnamefont {Demler}}, \ and\ \bibinfo
  {author} {\bibfnamefont {N.}~\bibnamefont {Goldman}},\ }\bibfield  {title}
  {\enquote {\bibinfo {title} {Parametric instability rates in periodically
  driven band systems},}\ }\href@noop {} {\bibfield  {journal} {\bibinfo
  {journal} {\Jprx}\ }\textbf {\bibinfo {volume} {7}},\ \bibinfo {pages}
  {021015} (\bibinfo {year} {2017})}\BibitemShut {NoStop}%
\bibitem [{\citenamefont {Modugno}\ \emph {et~al.}(2004)\citenamefont
  {Modugno}, \citenamefont {Tozzo},\ and\ \citenamefont {F.}}]{modugno2004}%
  \BibitemOpen
  \bibfield  {author} {\bibinfo {author} {\bibfnamefont {M.}~\bibnamefont
  {Modugno}}, \bibinfo {author} {\bibfnamefont {C.}~\bibnamefont {Tozzo}}, \
  and\ \bibinfo {author} {\bibfnamefont {Dalfovo}\ \bibnamefont {F.}},\
  }\bibfield  {title} {\enquote {\bibinfo {title} {Role of transverse
  excitations in the instability of bose-einstein condensates moving in optical
  lattices},}\ }\href@noop {} {\bibfield  {journal} {\bibinfo  {journal}
  {\Jpra}\ }\textbf {\bibinfo {volume} {70}},\ \bibinfo {pages} {043625}
  (\bibinfo {year} {2004})}\BibitemShut {NoStop}%
\bibitem [{\citenamefont {Kramer}\ \emph {et~al.}(2005)\citenamefont {Kramer},
  \citenamefont {Tozzo},\ and\ \citenamefont {Dalfovo}}]{Kramer:2005}%
  \BibitemOpen
  \bibfield  {author} {\bibinfo {author} {\bibfnamefont {M}~\bibnamefont
  {Kramer}}, \bibinfo {author} {\bibfnamefont {C}~\bibnamefont {Tozzo}}, \ and\
  \bibinfo {author} {\bibfnamefont {F}~\bibnamefont {Dalfovo}},\ }\bibfield
  {title} {\enquote {\bibinfo {title} {{Parametric excitation of a
  Bose-Einstein condensate in a one-dimensional optical lattice}},}\
  }\href@noop {} {\bibfield  {journal} {\bibinfo  {journal} {Physical Review
  A}\ }\textbf {\bibinfo {volume} {71}},\ \bibinfo {pages} {061602} (\bibinfo
  {year} {2005})}\BibitemShut {NoStop}%
\bibitem [{\citenamefont {Tozzo}\ \emph {et~al.}(2005)\citenamefont {Tozzo},
  \citenamefont {Kramer},\ and\ \citenamefont {Dalfovo}}]{tozzo2005}%
  \BibitemOpen
  \bibfield  {author} {\bibinfo {author} {\bibfnamefont {C.}~\bibnamefont
  {Tozzo}}, \bibinfo {author} {\bibfnamefont {M.}~\bibnamefont {Kramer}}, \
  and\ \bibinfo {author} {\bibfnamefont {F.}~\bibnamefont {Dalfovo}},\
  }\bibfield  {title} {\enquote {\bibinfo {title} {Stability diagram and growth
  rate of parametric resonances in bose-einstein condensates in one-dimensional
  optical lattices},}\ }\href@noop {} {\bibfield  {journal} {\bibinfo
  {journal} {\Jpra}\ }\textbf {\bibinfo {volume} {72}},\ \bibinfo {pages}
  {023613} (\bibinfo {year} {2005})}\BibitemShut {NoStop}%
\bibitem [{\citenamefont {Creffield}(2009)}]{creffield2009}%
  \BibitemOpen
  \bibfield  {author} {\bibinfo {author} {\bibfnamefont {C.~E.}\ \bibnamefont
  {Creffield}},\ }\bibfield  {title} {\enquote {\bibinfo {title} {Instability
  and control of a periodically-driven bose-einstein condensate},}\ }\href@noop
  {} {\bibfield  {journal} {\bibinfo  {journal} {\Jpra}\ }\textbf {\bibinfo
  {volume} {79}},\ \bibinfo {pages} {063612} (\bibinfo {year}
  {2009})}\BibitemShut {NoStop}%
\bibitem [{\citenamefont {Bukov}\ \emph
  {et~al.}(2015{\natexlab{b}})\citenamefont {Bukov}, \citenamefont
  {Gopalakrishnan}, \citenamefont {Knap},\ and\ \citenamefont
  {E.}}]{bukov2015}%
  \BibitemOpen
  \bibfield  {author} {\bibinfo {author} {\bibfnamefont {M.}~\bibnamefont
  {Bukov}}, \bibinfo {author} {\bibfnamefont {S.}~\bibnamefont
  {Gopalakrishnan}}, \bibinfo {author} {\bibfnamefont {M.}~\bibnamefont
  {Knap}}, \ and\ \bibinfo {author} {\bibfnamefont {Demler}\ \bibnamefont
  {E.}},\ }\bibfield  {title} {\enquote {\bibinfo {title} {Prethermal floquet
  steady states and instabilities in the periodically driven, weakly
  interacting bose-hubbard model},}\ }\href@noop {} {\bibfield  {journal}
  {\bibinfo  {journal} {\Jprl}\ }\textbf {\bibinfo {volume} {115}},\ \bibinfo
  {pages} {205301} (\bibinfo {year} {2015}{\natexlab{b}})}\BibitemShut
  {NoStop}%
\bibitem [{\citenamefont {Martin}\ \emph {et~al.}(2017)\citenamefont {Martin},
  \citenamefont {Georgeot}, \citenamefont {Guéry-Odelin},\ and\ \citenamefont
  {Shepelyansky}}]{martin2017}%
  \BibitemOpen
  \bibfield  {author} {\bibinfo {author} {\bibfnamefont {J.}~\bibnamefont
  {Martin}}, \bibinfo {author} {\bibfnamefont {B.}~\bibnamefont {Georgeot}},
  \bibinfo {author} {\bibfnamefont {D.}~\bibnamefont {Guéry-Odelin}}, \ and\
  \bibinfo {author} {\bibfnamefont {D.~L.}\ \bibnamefont {Shepelyansky}},\
  }\href@noop {} {\enquote {\bibinfo {title} {Kapitza stabilization of a
  repulsive bose-einstein condensate in an oscillating optical lattice},}\ }
  (\bibinfo {year} {2017}),\ \bibinfo {note} {{a}r{X}iv:1709.07792}\BibitemShut
  {NoStop}%
\bibitem [{\citenamefont {Michon}\ \emph {et~al.}(2017)\citenamefont {Michon},
  \citenamefont {Cabrera-Gutierrez}, \citenamefont {Fortun}, \citenamefont
  {Berger}, \citenamefont {Arnal}, \citenamefont {Brunaud}, \citenamefont
  {Billy}, \citenamefont {Petitjean}, \citenamefont {Schlagheck},\ and\
  \citenamefont {Guery-Odelin}}]{michon2017}%
  \BibitemOpen
  \bibfield  {author} {\bibinfo {author} {\bibfnamefont {E.}~\bibnamefont
  {Michon}}, \bibinfo {author} {\bibfnamefont {C.}~\bibnamefont
  {Cabrera-Gutierrez}}, \bibinfo {author} {\bibfnamefont {A.}~\bibnamefont
  {Fortun}}, \bibinfo {author} {\bibfnamefont {M.}~\bibnamefont {Berger}},
  \bibinfo {author} {\bibfnamefont {M.}~\bibnamefont {Arnal}}, \bibinfo
  {author} {\bibfnamefont {V.}~\bibnamefont {Brunaud}}, \bibinfo {author}
  {\bibfnamefont {J.}~\bibnamefont {Billy}}, \bibinfo {author} {\bibfnamefont
  {C.}~\bibnamefont {Petitjean}}, \bibinfo {author} {\bibfnamefont
  {P.}~\bibnamefont {Schlagheck}}, \ and\ \bibinfo {author} {\bibfnamefont
  {D.}~\bibnamefont {Guery-Odelin}},\ }\href@noop {} {\enquote {\bibinfo
  {title} {Out-of-equilibrium dynamics of a bose einstein condensate in a
  periodically driven band system},}\ } (\bibinfo {year} {2017}),\ \bibinfo
  {note} {arXiv:1707.06092}\BibitemShut {NoStop}%
\bibitem [{\citenamefont {Jaskula}\ \emph {et~al.}(2012)\citenamefont
  {Jaskula}, \citenamefont {Partridge}, \citenamefont {Bonneau}, \citenamefont
  {Lopes}, \citenamefont {Ruaudel}, \citenamefont {Boiron},\ and\ \citenamefont
  {Westbrook}}]{jaskula2012acoustic}%
  \BibitemOpen
  \bibfield  {author} {\bibinfo {author} {\bibfnamefont {J-C}\ \bibnamefont
  {Jaskula}}, \bibinfo {author} {\bibfnamefont {Guthrie~B}\ \bibnamefont
  {Partridge}}, \bibinfo {author} {\bibfnamefont {Marie}\ \bibnamefont
  {Bonneau}}, \bibinfo {author} {\bibfnamefont {Rapha{\"e}l}\ \bibnamefont
  {Lopes}}, \bibinfo {author} {\bibfnamefont {Josselin}\ \bibnamefont
  {Ruaudel}}, \bibinfo {author} {\bibfnamefont {Denis}\ \bibnamefont {Boiron}},
  \ and\ \bibinfo {author} {\bibfnamefont {Christoph~I}\ \bibnamefont
  {Westbrook}},\ }\bibfield  {title} {\enquote {\bibinfo {title} {Acoustic
  analog to the dynamical casimir effect in a bose-einstein condensate},}\
  }\href@noop {} {\bibfield  {journal} {\bibinfo  {journal} {Physical Review
  Letters}\ }\textbf {\bibinfo {volume} {109}},\ \bibinfo {pages} {220401}
  (\bibinfo {year} {2012})}\BibitemShut {NoStop}%
\bibitem [{\citenamefont {Bilitewski}\ and\ \citenamefont
  {Cooper}(2015{\natexlab{a}})}]{bilitewski_14}%
  \BibitemOpen
  \bibfield  {author} {\bibinfo {author} {\bibfnamefont {T.}~\bibnamefont
  {Bilitewski}}\ and\ \bibinfo {author} {\bibfnamefont {N.~R.}\ \bibnamefont
  {Cooper}},\ }\bibfield  {title} {\enquote {\bibinfo {title} {Population
  dynamics in a floquet realization of the harper-hofstadter hamiltonian},}\
  }\href@noop {} {\bibfield  {journal} {\bibinfo  {journal} {Phys. Rev. A}\
  }\textbf {\bibinfo {volume} {91}},\ \bibinfo {pages} {063611} (\bibinfo
  {year} {2015}{\natexlab{a}})}\BibitemShut {NoStop}%
\bibitem [{\citenamefont {Bilitewski}\ and\ \citenamefont
  {Cooper}(2015{\natexlab{b}})}]{bilitewski2015}%
  \BibitemOpen
  \bibfield  {author} {\bibinfo {author} {\bibfnamefont {T.}~\bibnamefont
  {Bilitewski}}\ and\ \bibinfo {author} {\bibfnamefont {N.~R.}\ \bibnamefont
  {Cooper}},\ }\bibfield  {title} {\enquote {\bibinfo {title} {Scattering
  theory for floquet-bloch states},}\ }\href@noop {} {\bibfield  {journal}
  {\bibinfo  {journal} {\Jpra}\ }\textbf {\bibinfo {volume} {91}},\ \bibinfo
  {pages} {033601} (\bibinfo {year} {2015}{\natexlab{b}})}\BibitemShut
  {NoStop}%
\bibitem [{\citenamefont {Weinberg}\ \emph {et~al.}(2015)\citenamefont
  {Weinberg}, \citenamefont {{\"O}lschl{\"a}ger}, \citenamefont {Str{\"a}ter},
  \citenamefont {Prelle}, \citenamefont {Eckardt}, \citenamefont {Sengstock},\
  and\ \citenamefont {Simonet}}]{weinberg2015}%
  \BibitemOpen
  \bibfield  {author} {\bibinfo {author} {\bibfnamefont {M.}~\bibnamefont
  {Weinberg}}, \bibinfo {author} {\bibfnamefont {C.}~\bibnamefont
  {{\"O}lschl{\"a}ger}}, \bibinfo {author} {\bibfnamefont {C.}~\bibnamefont
  {Str{\"a}ter}}, \bibinfo {author} {\bibfnamefont {S.}~\bibnamefont {Prelle}},
  \bibinfo {author} {\bibfnamefont {A.}~\bibnamefont {Eckardt}}, \bibinfo
  {author} {\bibfnamefont {K.}~\bibnamefont {Sengstock}}, \ and\ \bibinfo
  {author} {\bibfnamefont {J.}~\bibnamefont {Simonet}},\ }\bibfield  {title}
  {\enquote {\bibinfo {title} {Multiphoton interband excitations of quantum
  gases in driven optical lattices},}\ }\href@noop {} {\bibfield  {journal}
  {\bibinfo  {journal} {\Jpra}\ }\textbf {\bibinfo {volume} {92}},\ \bibinfo
  {pages} {043621} (\bibinfo {year} {2015})}\BibitemShut {NoStop}%
\bibitem [{\citenamefont {Sträter}\ and\ \citenamefont
  {Eckardt}(2016)}]{strater2016}%
  \BibitemOpen
  \bibfield  {author} {\bibinfo {author} {\bibfnamefont {C.}~\bibnamefont
  {Sträter}}\ and\ \bibinfo {author} {\bibfnamefont {A.}~\bibnamefont
  {Eckardt}},\ }\href@noop {} {\enquote {\bibinfo {title} {Interband heating
  processes in a periodically driven optical lattice},}\ } (\bibinfo {year}
  {2016}),\ \bibinfo {note} {arXiv:1604.00850}\BibitemShut {NoStop}%
\bibitem [{\citenamefont {{Quelle}}\ and\ \citenamefont {{Morais
  Smith}}(2017)}]{Quelle2017}%
  \BibitemOpen
  \bibfield  {author} {\bibinfo {author} {\bibfnamefont {A.}~\bibnamefont
  {{Quelle}}}\ and\ \bibinfo {author} {\bibfnamefont {C.}~\bibnamefont {{Morais
  Smith}}},\ }\bibfield  {title} {\enquote {\bibinfo {title} {{Resonances in a
  periodically driven bosonic system}},}\ }\href@noop {} {\bibfield  {journal}
  {\bibinfo  {journal} {ArXiv e-prints}\ } (\bibinfo {year} {2017})},\ \Eprint
  {http://arxiv.org/abs/1710.09680} {arXiv:1710.09680 [cond-mat.stat-mech]}
  \BibitemShut {NoStop}%
\bibitem [{\citenamefont {Logan W.~Clark}(2017)}]{clark2017}%
  \BibitemOpen
  \bibfield  {author} {\bibinfo {author} {\bibfnamefont {Lei Feng Cheng~Chin}\
  \bibnamefont {Logan W.~Clark}, \bibfnamefont {Anita~Gaj}},\ }\href@noop {}
  {\enquote {\bibinfo {title} {Collective emission of matter-wave jets from
  driven bose-einstein condensates},}\ } (\bibinfo {year} {2017}),\ \bibinfo
  {note} {{a}r{X}iv:1706.05560}\BibitemShut {NoStop}%
\bibitem [{\citenamefont {Salerno}\ \emph {et~al.}(2016)\citenamefont
  {Salerno}, \citenamefont {Ozawa}, \citenamefont {Price},\ and\ \citenamefont
  {Carusotto}}]{salerno2016}%
  \BibitemOpen
  \bibfield  {author} {\bibinfo {author} {\bibfnamefont {G.}~\bibnamefont
  {Salerno}}, \bibinfo {author} {\bibfnamefont {T.}~\bibnamefont {Ozawa}},
  \bibinfo {author} {\bibfnamefont {H.~M.}\ \bibnamefont {Price}}, \ and\
  \bibinfo {author} {\bibfnamefont {I.}~\bibnamefont {Carusotto}},\ }\bibfield
  {title} {\enquote {\bibinfo {title} {Floquet topological system based on
  frequency-modulated classical coupled harmonic oscillators},}\ }\href@noop {}
  {\bibfield  {journal} {\bibinfo  {journal} {\Jprb}\ }\textbf {\bibinfo
  {volume} {93}},\ \bibinfo {pages} {085105} (\bibinfo {year}
  {2016})}\BibitemShut {NoStop}%
\bibitem [{\citenamefont {Weinberg}\ \emph {et~al.}(2016)\citenamefont
  {Weinberg}, \citenamefont {Bukov}, \citenamefont {Vajna}, \citenamefont
  {D'Alessio}, \citenamefont {Polkovnikov},\ and\ \citenamefont
  {Kolodrubetz}}]{pweinberg_15}%
  \BibitemOpen
  \bibfield  {author} {\bibinfo {author} {\bibfnamefont {P.}~\bibnamefont
  {Weinberg}}, \bibinfo {author} {\bibfnamefont {M.}~\bibnamefont {Bukov}},
  \bibinfo {author} {\bibfnamefont {S.}~\bibnamefont {Vajna}}, \bibinfo
  {author} {\bibfnamefont {L.}~\bibnamefont {D'Alessio}}, \bibinfo {author}
  {\bibfnamefont {A.}~\bibnamefont {Polkovnikov}}, \ and\ \bibinfo {author}
  {\bibfnamefont {M.}~\bibnamefont {Kolodrubetz}},\ }\href@noop {} {\enquote
  {\bibinfo {title} {Adiabatic perturbation theory and geometry of
  periodically-driven systems},}\ } (\bibinfo {year} {2016}),\ \bibinfo {note}
  {arxiv:1606.02229}\BibitemShut {NoStop}%
\bibitem [{\citenamefont {Novicenko}\ \emph {et~al.}(2016)\citenamefont
  {Novicenko}, \citenamefont {Anisimovas},\ and\ \citenamefont
  {Juzeliunas}}]{novicenko2016}%
  \BibitemOpen
  \bibfield  {author} {\bibinfo {author} {\bibfnamefont {V.}~\bibnamefont
  {Novicenko}}, \bibinfo {author} {\bibfnamefont {E.}~\bibnamefont
  {Anisimovas}}, \ and\ \bibinfo {author} {\bibfnamefont {G.}~\bibnamefont
  {Juzeliunas}},\ }\bibfield  {title} {\enquote {\bibinfo {title} {Floquet
  analysis of a quantum system with modulated periodic driving},}\ }\href@noop
  {} {\bibfield  {journal} {\bibinfo  {journal} {arxiv:1608.08420}\ } (\bibinfo
  {year} {2016})}\BibitemShut {NoStop}%
\bibitem [{Note1()}]{Note1}%
  \BibitemOpen
  \bibinfo {note} {The factor of $2$ stems from the fact that $\Gamma $ is
  defined in amplitude while physical observables involve squared moduli of
  wavefunctions.}\BibitemShut {Stop}%
\bibitem [{\citenamefont {Choudhury}\ and\ \citenamefont
  {Mueller}(2015)}]{choudhury2015a}%
  \BibitemOpen
  \bibfield  {author} {\bibinfo {author} {\bibfnamefont {S.}~\bibnamefont
  {Choudhury}}\ and\ \bibinfo {author} {\bibfnamefont {E.}~\bibnamefont
  {Mueller}},\ }\bibfield  {title} {\enquote {\bibinfo {title} {Transverse
  collisional instabilities of a bose-einstein condensate in a driven
  one-dimensional lattice},}\ }\href@noop {} {\bibfield  {journal} {\bibinfo
  {journal} {\Jpra}\ }\textbf {\bibinfo {volume} {91}},\ \bibinfo {pages}
  {023624} (\bibinfo {year} {2015})}\BibitemShut {NoStop}%
\bibitem [{\citenamefont {Landau}\ and\ \citenamefont
  {Lifshits}(1969)}]{landau1969}%
  \BibitemOpen
  \bibfield  {author} {\bibinfo {author} {\bibfnamefont {L.}~\bibnamefont
  {Landau}}\ and\ \bibinfo {author} {\bibfnamefont {E.}~\bibnamefont
  {Lifshits}},\ }\href@noop {} {\emph {\bibinfo {title} {Theoretical Physics -
  Vol. 1 (mechanics)}}}\ (\bibinfo {year} {1969})\BibitemShut {NoStop}%
\bibitem [{\citenamefont {Stamper-Kurn}\ and\ \citenamefont
  {Ketterle}(2001)}]{stamper2001spinor}%
  \BibitemOpen
  \bibfield  {author} {\bibinfo {author} {\bibfnamefont {Dan~M}\ \bibnamefont
  {Stamper-Kurn}}\ and\ \bibinfo {author} {\bibfnamefont {Wolfgang}\
  \bibnamefont {Ketterle}},\ }\bibfield  {title} {\enquote {\bibinfo {title}
  {Spinor condensates and light scattering from bose-einstein condensates},}\
  }in\ \href@noop {} {\emph {\bibinfo {booktitle} {Coherent atomic matter
  waves}}}\ (\bibinfo  {publisher} {Springer},\ \bibinfo {year} {2001})\ pp.\
  \bibinfo {pages} {139--217}\BibitemShut {NoStop}%
\bibitem [{Note2()}]{Note2}%
  \BibitemOpen
  \bibinfo {note} {The global shift of $\pi /2$ comes from the fact that
  excitation momenta are defined with respect to the ground-state, which is a
  $\pi /2$ state in the present case.}\BibitemShut {Stop}%
\bibitem [{\citenamefont {Goldstone}\ \emph {et~al.}(1962)\citenamefont
  {Goldstone}, \citenamefont {Salam},\ and\ \citenamefont
  {Weinberg}}]{goldstone1962}%
  \BibitemOpen
  \bibfield  {author} {\bibinfo {author} {\bibfnamefont {J.}~\bibnamefont
  {Goldstone}}, \bibinfo {author} {\bibfnamefont {A.}~\bibnamefont {Salam}}, \
  and\ \bibinfo {author} {\bibfnamefont {S.}~\bibnamefont {Weinberg}},\
  }\bibfield  {title} {\enquote {\bibinfo {title} {Broken symmetries},}\
  }\href@noop {} {\bibfield  {journal} {\bibinfo  {journal} {\Jpr}\ ,\ \bibinfo
  {pages} {965–970}} (\bibinfo {year} {1962})}\BibitemShut {NoStop}%
\bibitem [{\citenamefont {Powell}\ \emph {et~al.}(2011)\citenamefont {Powell},
  \citenamefont {Barnett}, \citenamefont {Sensarma},\ and\ \citenamefont
  {Das~Sarma}}]{powell2011}%
  \BibitemOpen
  \bibfield  {author} {\bibinfo {author} {\bibfnamefont {S.}~\bibnamefont
  {Powell}}, \bibinfo {author} {\bibfnamefont {R.}~\bibnamefont {Barnett}},
  \bibinfo {author} {\bibfnamefont {R.}~\bibnamefont {Sensarma}}, \ and\
  \bibinfo {author} {\bibfnamefont {S.}~\bibnamefont {Das~Sarma}},\ }\bibfield
  {title} {\enquote {\bibinfo {title} {Bogoliubov theory of interacting bosons
  on a lattice in a synthetic magnetic field},}\ }\href@noop {} {\bibfield
  {journal} {\bibinfo  {journal} {\Jpra}\ }\textbf {\bibinfo {volume} {83}},\
  \bibinfo {pages} {013612} (\bibinfo {year} {2011})}\BibitemShut {NoStop}%
\bibitem [{Note3()}]{Note3}%
  \BibitemOpen
  \bibinfo {note} {Let us stress that the effective Hamiltonian in Eq.~(\ref
  {eq:Heff2D}) obtained from Eq.~(\ref {eq:H2D}), does not correspond to the
  Landau gauge, hence the difference compared to the more common version of the
  Harper-Hofstadter model whose spectrum is $E_{\protect \mathbf {p}}^{\pm
  }=\pm 2 \protect \sqrt {{J_{\protect \mathrm {eff}}^x}^2 \protect \qopname
  \relax o{cos}^2 p_x + {J_{\protect \mathrm {eff}}^y}^2 \protect \qopname
  \relax o{cos}^2 p_y}$ and ground-state in $p_x=p_y=0$.}\BibitemShut {Stop}%
\bibitem [{Note4()}]{Note4}%
  \BibitemOpen
  \bibinfo {note} {Interestingly however, we find that this term can play a
  role for higher values of $l$, with non trivial effects arising from the
  non-commutation between the diagonal $\protect \mathaccentV
  {hat}05EW_\protect \mathbf {q}(t)$ and the off-diagonal matrix in Eqs.~\ref
  {eq:appBdGEPO}.}\BibitemShut {Stop}%
\end{thebibliography}
\end{document}